\documentclass[a4paper,fleqn,usenatbib]{mnras}

\usepackage{amsmath}	
\usepackage{mathptmx}
\usepackage{txfonts}

\usepackage[T1]{fontenc}
\usepackage{ae,aecompl}

\usepackage[final]{graphicx}	% Including figure files
\newcommand\HI{H\,{\sc i} }%

\title[\HI replenishment from the cosmic web]{Evidence for \HI replenishment in massive galaxies through gas accretion from the cosmic web}

\author[D. Kleiner et al.]
	{Dane Kleiner,$^{1, 2, 3}$\thanks{E-mail: dane.kleiner@monash.edu}
	Kevin A. Pimbblet,$^{1, 3}$
	D. Heath Jones,$^{1, 4, 5}$
	B\"arbel S. Koribalski$^{2}$	
	\newauthor and Paolo Serra$^{2, 6}$
\\
$^{1}$ School of Physics and Astronomy, Monash University, Melbourne, VIC 3800, Australia\\
$^{2}$ CSIRO Astronomy \& Space Science, Australia Telescope National Facility, PO Box 76, Epping, NSW 1710, Australia\\
$^{3}$ E. A. Milne Centre for Astrophysics, University of Hull, Cottingham Road, Kingston-upon-Hull HU6 7RX, UK\\
$^{4}$ Department of Physics and Astronomy, Macquarie University, Sydney, NSW 2109, Australia\\
$^{5}$ English Language and Foundation Studies Centre, University of Newcastle, Callaghan, NSW 2308, Australia\\
$^{6}$ INAF -- Osservatorio Astronomico di Cagliari, Via della Scienza 5, I-09047 Selargius (CA), Italy\\
}

\date{Accepted December 19, 2016. Received November 16, 2016; in original form July 13, 2016.}

\pubyear{2016}

\begin{document}
\label{firstpage}
\pagerange{\pageref{firstpage}--\pageref{lastpage}}
\maketitle

\begin{abstract}
%It should be a single paragraph not more than 250 words (200 words for Letters).
We examine the H\,{\sc i}-to-stellar mass ratio (\HI fraction) for galaxies near filament backbones within the nearby Universe ($d <$ 181 Mpc). This work uses the 6 degree Field Galaxy Survey and the Discrete Persistent Structures Extractor to define the filamentary structure of the local cosmic web. \HI spectral stacking of \HI Parkes all sky survey observations yields the \HI fraction for filament galaxies and a field control sample. The \HI fraction is measured for different stellar masses and fifth nearest neighbour projected densities ($\Sigma_{5}$) to disentangle what influences cold gas in galaxies. For galaxies with stellar masses log($M_{\star}$) $<$ 11 M$_{\odot}$ in projected densities 0 $\leq$ $\Sigma_{5}$ $<$ 3 galaxies Mpc$^{-2}$, all \HI fractions of galaxies near filaments are statistically indistinguishable from the control sample. Galaxies with stellar masses log($M_{\star}$) $\geq$ 11 M$_{\odot}$ have a systematically higher \HI fraction near filaments than the control sample. The greatest difference is 0.75 dex, which is 5.5$\sigma$ difference at mean projected densities of 1.45 galaxies Mpc$^{-2}$. We suggest that this is evidence for massive galaxies accreting cold gas from the intrafilament medium that can replenish some \HI gas. This supports cold mode accretion where filament galaxies with a large gravitational potential can draw gas from the large-scale structure.

%No references should appear in the abstract.

\end{abstract}


%%%%%%%%%%%%%%%%%%%%%%%%%%%%%%%%%%%%%%%%%%%%%%%%%%

%%%%%%%%%%%%%%%%% BODY OF PAPER %%%%%%%%%%%%%%%%%%

\section{Introduction}

The distribution of galaxies in the Universe forms a cosmic web \citep{Bond1996} consisting of clusters, filaments and voids. Tightly packed groups and clusters are situated at web nodes and interconnected by a tenuous network of galaxy filaments delineating the underdense voids. These features were observed in early galaxy redshift surveys \citep[e.g.~CfA; ][]{Huchra1983, Lapparent1986} and are clear in more recent galaxy redshift surveys such as the 2 degree Field Galaxy Redshift Survey (2dFGRS), the Sloan Digital Sky Survey (SDSS), the 6 degree Field Galaxy Survey (6dFGS), the Galaxy And Mass Assembly survey (GAMA) and the 2MASS Redshift Survey \citep{Colless2001, Abazajian2009, Jones2009, Driver2011, Huchra2012}. Dark matter $N$-body simulations and baryonic hydrodynamical simulations also reproduce these structures \citep[e.g.][]{Springel2006, Popping2009, Angulo2012, Habib2012, Poole2015}. 

Although filaments, clusters and voids have been studied separately for a long time, a unified approach to all these structures have only been possible in the past few decades due to the availability of wide-field redshift surveys. The cosmic web spans many orders of magnitude where the boundaries between the features of the web are not clearly defined. Finding a reliable method that quantifies galaxy membership of different regions is a difficult task and many attempts have been presented \citep[e.g.][]{Pimbblet2005, Stoica2005, Novikov2006, Hahn2007, Platen2007, Neyrinck2008, Forero-Romero2009, Stoica2010, Aragon-Calvo2010b, Aragon-Calvo2010c, Sousbie2011a, Sousbie2011b, Smith2012, Leclercq2015}. Currently, no single method has been universally adopted for such classification.

Furthermore, galaxy redshift survey data are often sparse and incomplete. Despite this, attempts to disentangle the cosmic web from these surveys are wide ranging \citep[2dFGRS, SDSS, GAMA;][]{Pimbblet2004, Murphy2011,Tempel2014, Chen2015b, Alpaslan2014}.

40 per cent of the mass in the Universe resides within filaments \citep{Aragon-Calvo2010a},  which makes them an important aspect of galaxy evolution. While mass \citep[e.g.][]{Baldry2006, vonderLinden2010, Evoli2011, Pimbblet2012, Lemonias2013, Lin2016} and local galaxy density \citep[e.g.][]{Balogh2004, Kauffmann2004, Baldry2006, Poggianti2006, Park2007, Bamford2009} are primary drivers of galaxy evolution, the total baryonic mass residing in filaments is too large to ignore. More broadly, filaments can be classified as intermediate density environments that (predominantly) host small groups of galaxies. \citet{Chen2015a} show that the properties of filament galaxies adhere to the morphology--density relation of \citet{Dressler1980}. The moderately dense filament environment makes them suitable sites for galaxy pre-processing, where galaxies may be significantly transformed before entering a cluster's virial radius. Pre-processing can strip the gas from a galaxy and quench its star formation \citep[e.g.][]{Lewis2002, Gomez2003, Porter2008, Haines2011, Mahajan2013} and filaments are ideal environments to search for signatures of galaxy pre-processing. 

Previous attempts to determine if galaxies evolve differently in filaments are illuminating. A study by \citet{Porter2008} found a weak enhancement of star formation in the faint dwarf galaxies contained in filaments feeding a supercluster. \citet{Mahajan2012} show a similar result for the star formation rate of galaxies feeding clusters. However, \citet{Alpaslan2015} found that galaxy colour, morphology and stellar mass are driven by the stellar mass and local environment of galaxies and there was no observable influence that were driven by filaments. Later work by \citet{Alpaslan2016} found that the star formation of galaxies closer to the filament axis is slightly lower than those further away. This is consistent with \citet{Chen2015b}, who showed red high-mass galaxies are typically closer to filaments than blue low-mass galaxies. \citet{Martinez2016} find that galaxies in filaments have a lower specific star formation rate than galaxies in the field. Taken together, these studies suggest that galaxies in filaments evolve differently to galaxies in other large-scale environments. However, the evidence is tentative and more work is needed to confidently measure the potential second-order influence of the large-scale structure.

Numerical and hydrodynamic simulations show these filaments host a diffuse gaseous medium with a neutral hydrogen (H\,{\sc i}) column density of $\sim$10$^{16}$ cm$^{-2}$ at low redshift \citep[e.g.][]{Popping2009}. Current radio telescopes observe column densities down to $\sim$10$^{18}$ cm$^{-2}$ which is not sensitive enough to detect the proposed \HI in filaments. Lyman $\alpha$ emission from distant quasars can ``illuminate'' the gas in the cosmic web which has constrained the mass and morphology of gas in the cosmic web \citep[e.g.][]{Cantalupo2014}. However, these systems are rare and cannot be used to study the cosmic web as a large statistical sample. 

The interaction between galaxies and the gas reservoir in filaments is poorly constrained and where galaxies accrete their gas from is not well understood. A favourable scenario for gas accretion on to galaxies is `cold mode' accretion where the gaseous medium of filaments provides galaxies with a reservoir of gas to draw from \citet{Keres2005}. Recently, \citet{Aragon-Calvo2016} proposed the cosmic web detachment (CWD) model, which links the properties of galaxies in a natural cosmological framework. In this model, galaxies attached to the cosmic web are able to accrete cold gas from filaments and form stars. Non-linear interactions detach galaxies from the cosmic web, sever its cold gas supply and quench star formation. The CWD model successfully reproduces important features of galaxy evolution (such as the red fraction dependence on mass and environment) in $N$-body simulations. However, observational evidence supporting this model is currently lacking.

Understanding the \HI content of galaxies is fundamental to understanding galaxy evolution. A sufficiently dense supply of \HI can condense into massive molecular clouds that collapse and forms stars. Galaxies with an abundance of \HI have the ability to form new stars through this process and there are strong correlations between the \HI content of a galaxy and its star formation \citep{Doyle2006, Cortese2011, Fabello2011a, Catinella2013, Brown2015}. Galaxies devoid of \HI gas do not have the fuel to form new stars and \HI-deficient galaxies are correlated with passive galaxies \cite[e.g.][]{Catinella2010, Denes2014, Denes2016}. Another important aspect is that \HI extends to larger radius than stars in galaxies. Therefore, it is more easily perturbed during tidal interactions, and can exhibit a disturbed morphology for a long period of time after these events \citep[e.g.][]{Koribalski2004a, Holwerda2011}, making the \HI content of galaxies a powerful probe of their interaction with the environment.

The \HI Parkes all sky survey \citep[HIPASS;][]{Staveley-Smith1996,Barnes2001} detected \HI in 4315 galaxies in the southern sky out to a redshift of $z <$ 0.0423 \citep{Meyer2004, Zwaan2004} down to a column density of 4.0 $\times$ 10$^{18}$ cm$^{-2}$. HIPASS has played a significant part in understanding the role of \HI in galaxies \citep[e.g.][]{Meyer2004, Zwaan2004, Koribalski2004b, Zwaan2005, Doyle2005, Wong2006, Meyer2008, Wong2009, Popping2011, Allison2014} but is limited to only the brightest \HI galaxies. Since the \HI 21-cm emission line is weak, observing \HI in galaxies with low column densities requires long integration times to reach the desired sensitivity that is often impractical. The Square Kilometre Array (SKA) and Australian Square Kilometre Array Pathfinder (ASKAP) will be able to observe faint \HI galaxies with reasonable integration times. Survey such as the Widefield ASKAP L-band Legacy All-sky Blind surveY \citep[WALLABY;][]{Koribalski2012} and the Deep Investigation of Neutral Gas Origins (DINGO) will allow us to directly study the \HI in galaxies with unprecedented detail in larger volumes than any current survey. However WALLABY, DINGO and SKA surveys will not be available for years to come and we must rely on indirect methods of measuring \HI in galaxies.

In the interim, spectral stacking has proven to be a powerful technique for measuring \HI line strength. It measures the average \HI content of galaxies by co-adding spectra together \citep{Lah2007, Lah2009, Pen2009, Fabello2011a, Fabello2011b, Fabello2012, Delhaize2013, Gereb2013, Gereb2015,  Meyer2016}. The advantage of spectral stacking is improved signal-to-noise over an individual spectrum, revealing signatures that cannot be individually detected. As \HI in galaxies is sensitive to the dynamical history and immediate environment, the \HI mass of a galaxy can span orders of magnitude for a given stellar mass. The H\,{\sc i}-to-stellar mass ratio (\HI fraction) is a useful way of comparing how much \HI is present in galaxies with different stellar masses. There is a strong anticorrelation between \HI fraction and galaxy stellar mass while there is a weaker anti-correlation between \HI fraction and local galaxy density \citep[e.g.][]{Baldry2008, Catinella2010, Cortese2011, Brown2015}. By contrasting the \HI fraction of galaxies in filaments with that of galaxies far from filaments (at fixed stellar mass and local galaxy density), we can directly compare the availability of \HI gas as a function large-scale environment and test the CWD model.

In this paper, we select galaxies from the filament backbones of 6dFGS \citep{Jones2009} and measure the average \HI fraction from HIPASS \citep{Barnes2001} for different stellar masses and local galaxy densities. The 6dFGS is the most comprehensive galaxy redshift survey of the local Universe in the Southern hemisphere to date. There cosmic web is well-defined within $z <$ 0.1 making it ideal for studying the large-scale structure of the Universe and the galaxies that reside in it. This work can be used as a pilot study to define the cosmic web in the upcoming TAIPAN \citep{Hopkins2014} survey and predict the amount of \HI present in galaxies within the cosmic web for ASKAP and the SKA. 

Section \ref{data} presents the optical data and \HI data. In Section \ref{filaments}, we present our delineation of filaments in 6dFGS and the samples used in this study. Section \ref{stacking} is where we show the stacking technique used to measure the average \HI fraction of our samples. Section \ref{results} presents the results for the average \HI fraction our samples in different stellar mass and local density regimes. Section \ref{discussion} discusses our measured \HI fractions and its implication of galaxies evolving in the cosmic web. Finally, we summarize our conclusions in Section \ref{conclusions}. Throughout this paper, we assume the standard $\Lambda$ cold dark matter cosmological model with $\Omega_{\rm M}$ = 0.3, $\Lambda$ = 0.7 and H$_{0}$ = 70.4 km s$^{-1}$ Mpc$^{-1}$ \citep{Komatsu2011}. 

\section{Data}
\label{data}
In this work, we utilize two primary data sets. The first is the 6dFGS \citep{Jones2004, Jones2009}. We use the positions and redshifts of 6dFGS galaxies to determine the filamentary structure. From the 6dFGS catalogue, we use the SuperCOSMOS $B$--band photometry to test for unreliable \HI spectra (Section \ref{source_conf}) and we use the 2MASS $J$--band photometry to estimate stellar masses in 6dFGS. The second data set we use is the HIPASS \citep{Staveley-Smith1996, Barnes2001}. We extract the \HI spectra from HIPASS data cubes using the optical positions and redshifts which are then stacked for different samples. In the following sections, we describe each data set and how they relate to this work. 

\subsection{6dFGS optical catalogue}
The 6dFGS is a galaxy redshift and peculiar velocity survey conducted in the Southern Hemisphere beyond $\lvert b \rvert >$ 10$^{\circ}$ \citep{Jones2004, Jones2009}. It contains 125\,071 galaxy redshifts obtained with the 6 degree field fibre-fed multi-object spectrograph at the United Kingdom Schmidt Telescope at Siding Spring. The survey is near-complete at ($K$, $H$, $J$ , $r_{\rm F}$, $b_{\rm J}$) $\leq$ (12.65, 12.95, 13.75, 15.60, 16.75). The primary targets for 6dFGS are the $K$-band selected and secondary targets are selected from the $J$, $H$ $b_{\rm J}$ and $r_{\rm F}$ bands. The $J-$, $H-$ and $K$-band photometry is taken from the 2MASS extended source catalogue \citep{Jarrett2000} and the $b_{\rm J}$ and $r_{\rm F}$ photometry comes from the SuperCOSMOS catalogue \citep{Hambly2001a, Hambly2001b}. The 6dFGS volume is comparable to 2dFGRS \citep{Colless2001}, but a consequence of the huge sky coverage of 6dFGS is that it has a bright flux limit and therefore a relatively shallow survey depth compared to the other surveys. 

Each 6dFGS redshift is assigned a quality ($Q$) reflecting its reliability. The redshifts blunder rate for individual and pairs of 6dFGS galaxies is 1.2 and 2.3 per cent, respectively. Redshift blunders are defined as a mismatch of more than 330 km s$^{-1}$ (5$\sigma$) between a pair of redshift measurements of the same target. We refer the reader to \citep{Jones2004, Jones2009} for further technical details. 

For this analysis, we limit the 6dFGS input sample to redshift-space positions within the HIPASS volume ($z <$ 0.0423). Only galaxies with $Q$ = 3 (reliable) or 4 (high quality) have been used to define our cosmic structures and extract \HI spectra for subsequent stacking. As the HIPASS volume is less than the median redshift of 6dFGS ($z_{1/2}$  = 0.053), all galaxies within the 6dFGS catalogues are used irrespective of the subsample to which they belong. 

Stellar mass estimates are those used in \citet{Beutler2013}, which have been applied to our cosmology (H$_{0}$ = 70.4 km s$^{-1}$ Mpc$^{-1}$). The 6dFGS $J$--band magnitudes we use offer the lowest background noise making it the most reliable for stellar mess estimates and have been used in the calculations. Stellar population synthesis results from \citet{Bruzual1993} were used in conjunction with the modified Salpeter initial mass function \citep{Bell2001} to estimate the stellar masses. The typical uncertainty for the stellar mass estimates within this volume are log($M_{\star}$) $\pm$ 0.04 M$_{\odot}$.

\subsection{HIPASS radio data}
HIPASS is a blind \HI survey that covers the whole southern sky \citep{Barnes2001} and northern sky to $\delta$ $\leq$ +25$^{\circ}$ \citep{Wong2006}. The observations were carried out with the Australian Telescope National Facility's Parkes 64 m telescope using the 21 cm multibeam receiver \citep{Staveley-Smith1996}. The observed frequency range is 1362.5 -- 1426.5 MHz corresponding to a velocity range $-$1\,280 < c$z$ < 12\,700 km s$^{-1}$. The data are organized into 8$^{\circ}$ $\times$ 8$^{\circ}$ cubes with a gridded beam of $\sim$15.5 arcmin and 18 km s$^{-1}$ velocity resolution. Like other surveys that observe a large area of sky, it has a shallow survey depth ($z <$ 0.0423) with a rms of 13.3 mJy per 13 km s$^{-1}$ channel. These limitations result in only the brightest \HI galaxies being detected in HIPASS. As we are utilizing a spectral stacking technique in our analysis, it is not of concern if an individual galaxy has been detected above the noise. 

For this study, we use HIPASS data cubes to extract \HI spectra to conduct our stacking analysis. The positions and redshifts of 6dFGS galaxies are used to define the central co-ordinates in the HIPASS data cubes which we extract individual \HI spectra for galaxies of different samples. The spectra are stacked and we measure the \HI fraction for each sample.

\section{Delineating filaments from redshift data}
\label{filaments}
To define filaments of galaxies in redshift data, one must devise a method of connecting a series of points that have a higher density than a random distribution and determine if the result is a good representation of the large-scale structure in the Universe. This is inherently difficult as the length of filaments spans many orders of magnitude with complex geometry and poorly defined boundaries. Redshift data present additional challenges as the data can be sparse, incomplete and redshift space distortions elongate clusters in the $z$-direction, causing them to appear filamentary when they should not be classified as filaments. 

\subsection{Choice of algorithm}
\label{fil_algorithms}
Numerous techniques have been developed in an attempt to overcome the difficulties of reliability delineating filaments in galaxy surveys. These techniques are separated by a set of assumptions based on the physical, probabilistic and geometric treatment of filaments.  

Two physically motivated filament finders are the two-pass friends-of-friends algorithm and minimal spanning tree used by \citet{Murphy2011} and \citet{Alpaslan2014}. Both approaches operate on a group catalogue where redshift space distortions are suppressed. This is advantageous by ensuring clusters stretched in the $z$-direction do not get classified as filaments. A limitation of this approach is that mock catalogues that match the survey characteristics (i.e.~selection function and completeness) are required to calibrate the parameters that link galaxies in groups and groups to filaments. The linking parameters for extracting filaments in this technique are arbitrary in the sense that it is visually determined by what looks the best.

The probabilistic approach for delineating filaments uses galaxy positions to determine the likelihood of being a part of a long, thin structure (i.e.~a filament) in random data. \citet{Tempel2014} use the Bisous model, an object point process with interactions that assume galaxies are grouped inside small cylinders and connected aligned cylinders form filaments. \citet{Leclercq2015} use decision theory, a method that classifies cosmic web components by deciding which component maximises its quantitative profit for each galaxy position. Filaments have a unique structural type  component in decision theory and galaxy positions optimized by this component are chosen to be part of the filament. These techniques are a natural integration of probability theory that allows the classification of filaments directly from the galaxy positions without a group catalogue or density field. However, redshift space distortions are not suppressed and the success of these techniques relies on how well the priors match the underlying cosmic web. Currently, the optimal choice for modelling filaments is an open mathematical and data analysis problem. 

A popular technique for extracting filaments uses a geometric approach. Broadly speaking, a density field or equivalent (such as a tessellation) is created from galaxy positions and a mathematical formalization is computed over the field to compute the geometry and extract filaments. For example, the watershed technique transforms the density field into a series of valleys and ridges and filaments are viewed as ridges surrounded by multiple valleys \citep{Platen2007, Aragon-Calvo2010c}, the Hessian matrix computed from a density field identifies filaments as a specific set of eigenvalues \citep{Hahn2007, Cautun2013}, the subspace constrained mean shift algorithm identifies filaments as ridges above a critical density \citep{Chen2015b} and connected saddle points in Morse theory are used to delineate filaments from the Delaunay tessellation \citep{Sousbie2011a}. While filaments are viewed and extracted differently in each formalism, every technique using the geometric approach requires a measure of density and filaments are identified as thin ridges. The advantage of the geometric approach is that it does not make any assumptions about the underlying dark matter distribution of the Universe. The success of each technique relies on how well the mathematical formalism identify and separate the components of the cosmic web in data with varying densities. Even though there is no physical basis for the cosmic web to be described by the formalisms, all these techniques have shown excellent agreement with identifying filaments (and other cosmic web components). 

The geometric approaches \citep[such as][]{Aragon-Calvo2010c, Sousbie2011a, Cautun2013} have produced excellent filament delineation in $N$-body simulations. When searching for filaments in redshift data, incorrect distances measured from redshifts make it considerably harder to trace the coherent three-dimensional structure of filaments. This is a serious concern in the local Universe where peculiar velocities dominate redshift measurements of galaxies. \citet{Chen2015a} splits the galaxy positions into thin redshift slices, removing the redshift dependence of filaments. While the three-dimensional coherence of filaments is lost, the galaxy number density in each slice is sufficient to easily observe the large-scale structure of filaments at that distance. Hence, the contamination from peculiar velocity dominated redshifts is minimized. 

Given the above considerations, we have decided to identify filaments in 6dFGS using the Discrete Persistent Structures Extractor\footnote{\url{http://www2.iap.fr/users/sousbie/web/html/indexd41d.html}} \citep[DisPerSE;][]{Sousbie2011a,Sousbie2011b}. DisPerSE is freely available, easy to use and utilizes a geometric approach to identify coherent multiscale astrophysical structures such as voids, walls, filaments and clusters. While there are multiple geometric approaches that can be used to identify filaments, DisPerSE has the distinct advantage of operating directly on a tessellation opposed to a smoothed density field where structural information can be lost (e.g.~Watershed, Hessian matrix, subspace constrained mean shift algorithm). Additionally, DisPerSE is scale- and parameter-free unlike other geometric approaches that require smoothing lengths. DisPerSE can be applied to both $N$-body simulations and galaxy redshift surveys without the use of complex mocks, and \citet{Sousbie2011b} show excellent recovery of the cosmic web in both simulation and redshift surveys, even when the data is sparse and incomplete. 

\subsection{Filament backbones in 6dFGS}
\label{fil_6dFGS}
We summarize the main steps below and refer the reader to \citet{Sousbie2011a} for the theory and implementation of DisPerSE, \citet{Sousbie2011b} for the application of DisPerSE on astronomical data and appendix B of \citet{Sousbie2011b} for a succinct summary of the algorithm.

We adopt a similar method to \citet{Chen2015a} and split 6dFGS into thin redshift slices ($\delta$$z$ = 0.005) to run DisPerSE on two-dimensional data. While DisPerSE can be applied to three-dimensional data, we found poor agreement between the cosmic web and the filamentary skeleton returned by DisPerSE. See Appendix \ref{appen:disperse_compare} for a comparison and discussion on the application of DisPerSE in two and three dimensions. 

DisPerSE operates directly on the galaxy positions and we separate each slice into distinct samples above and below the Galactic plane ($\lvert b \rvert >$ 10$^{\circ}$) to ensure no spurious connections are made over this boundary. A density field is produced from the galaxy positions using the Delaunay tessellation. Discrete Morse theory \citep{Forman1998, Forman2002} is then applied on the tessellation, which computes the geometry and topology of the density field, enabling the filament backbones to be recovered. Fig.~\ref{fig:disperse} shows this procedure on two slices used in this analysis and the full sample can be viewed in Appendix \ref{appen:disperse}.

\begin{figure*}
\centering
\includegraphics[scale=.44, trim={5cm 0cm 7cm 0cm}, clip]{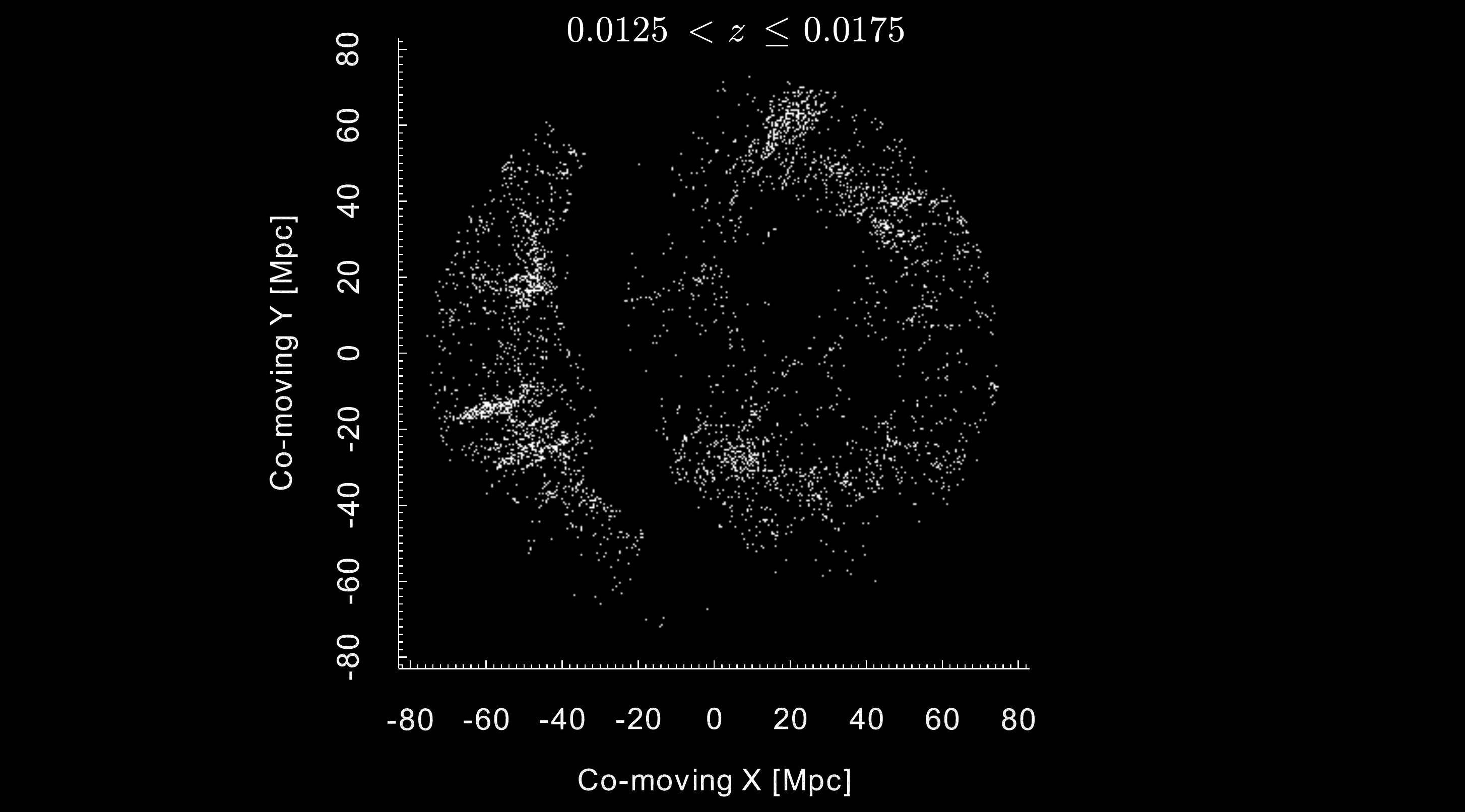}
\includegraphics[scale=.44, trim={5cm 0cm 7cm 0cm}, clip]{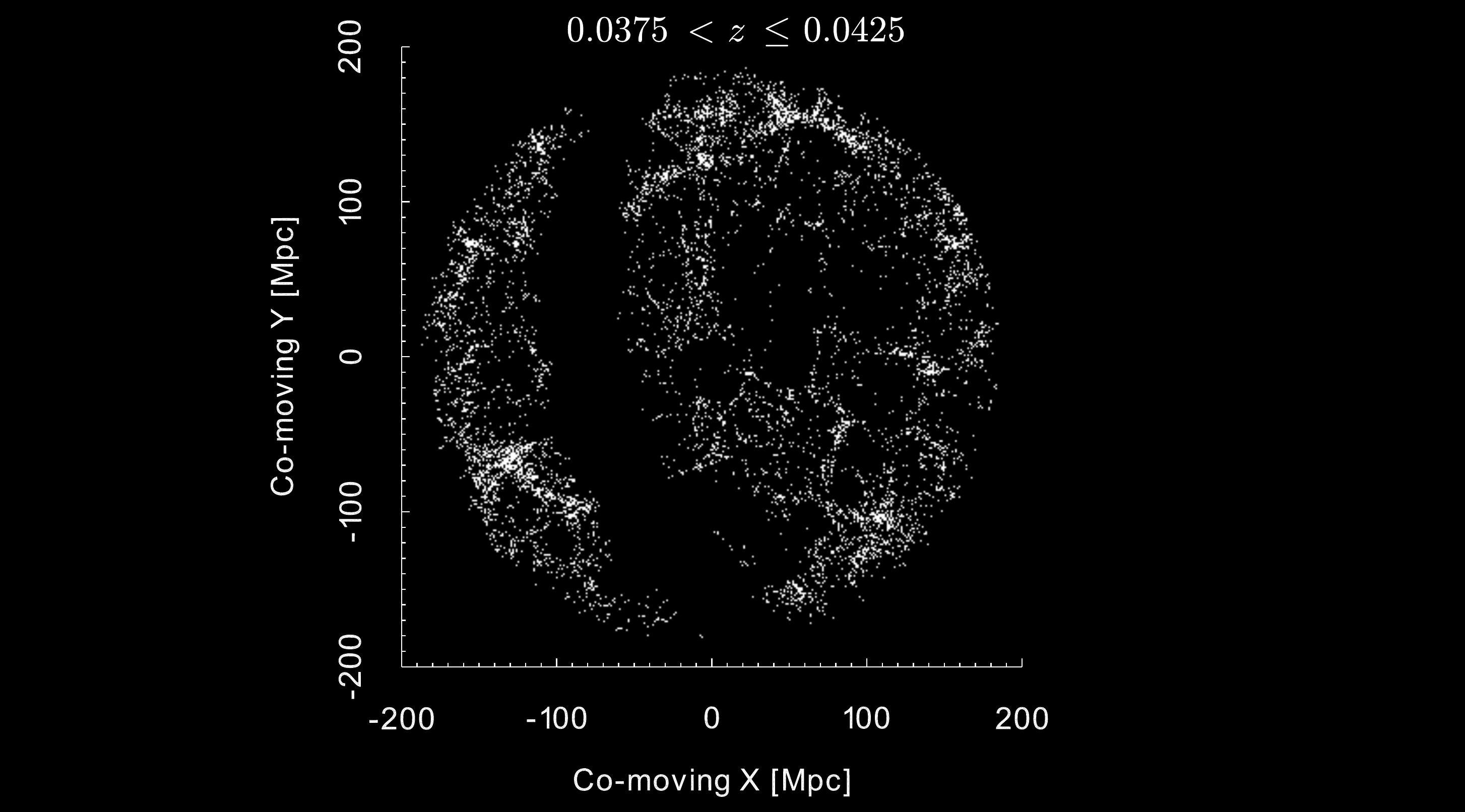}

\includegraphics[scale=.44, trim={5cm 0cm 7cm 0cm}, clip]{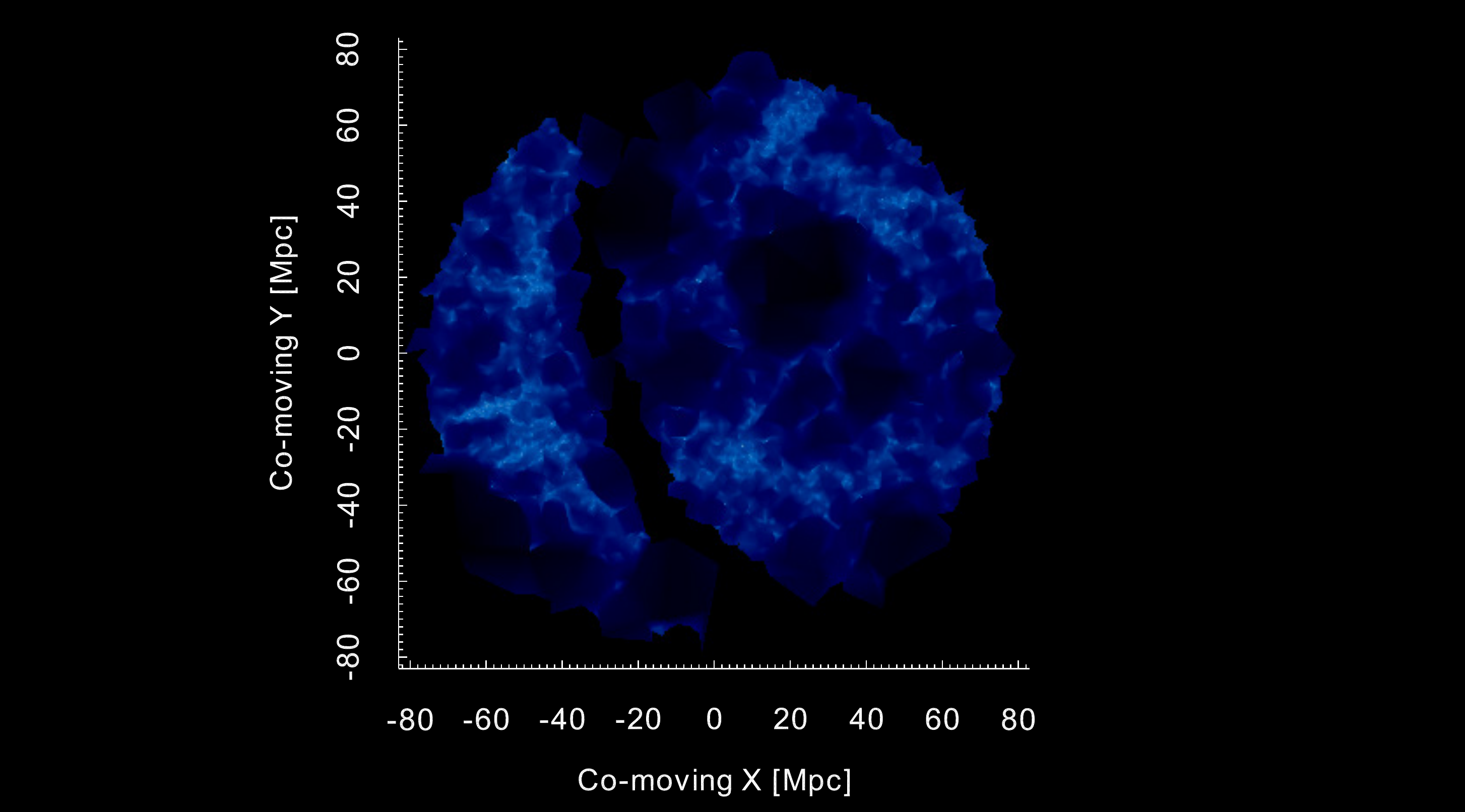}
\includegraphics[scale=.44, trim={5cm 0cm 7cm 0cm}, clip]{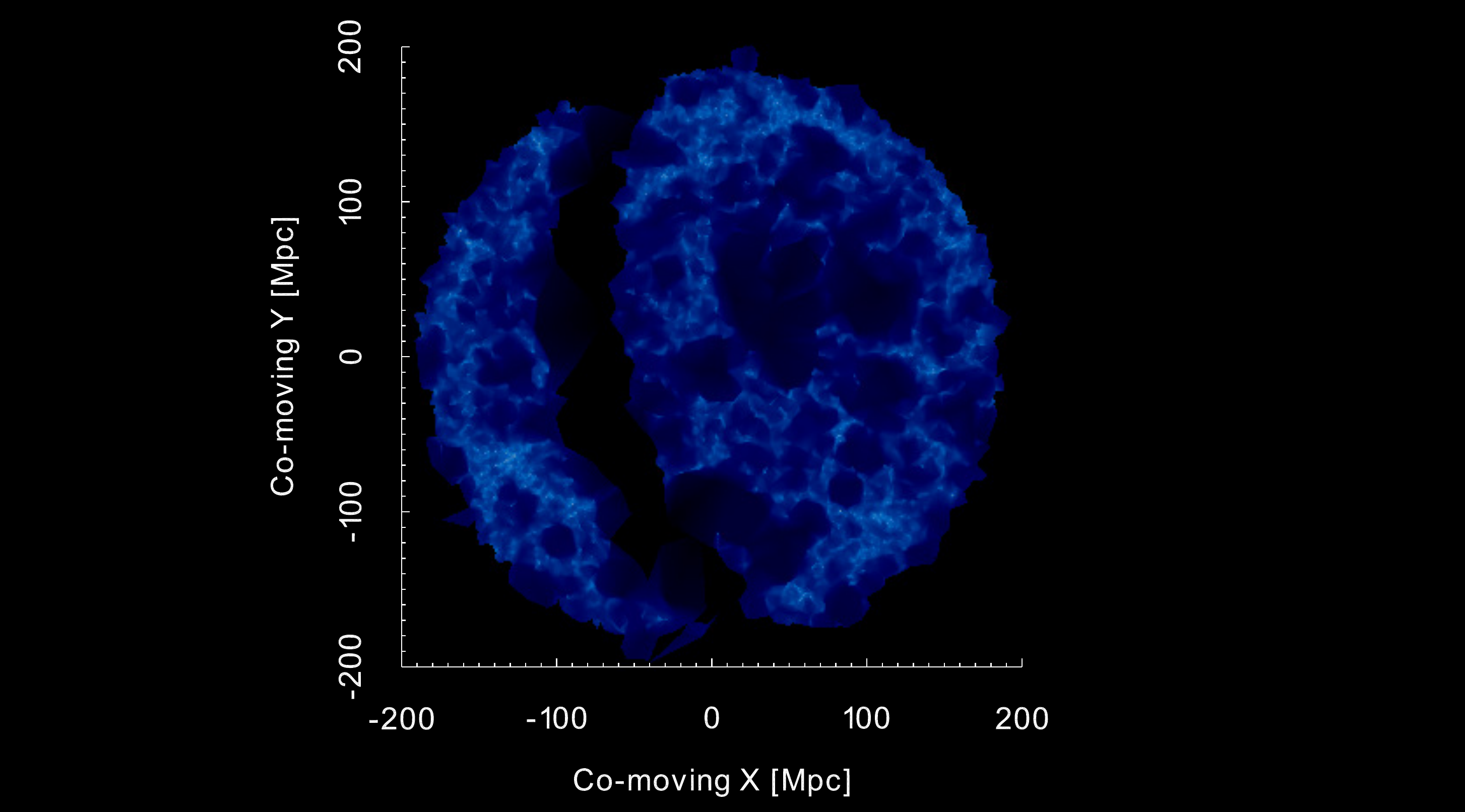}

\includegraphics[scale=.44, trim={5cm 0cm 7cm 0cm}, clip]{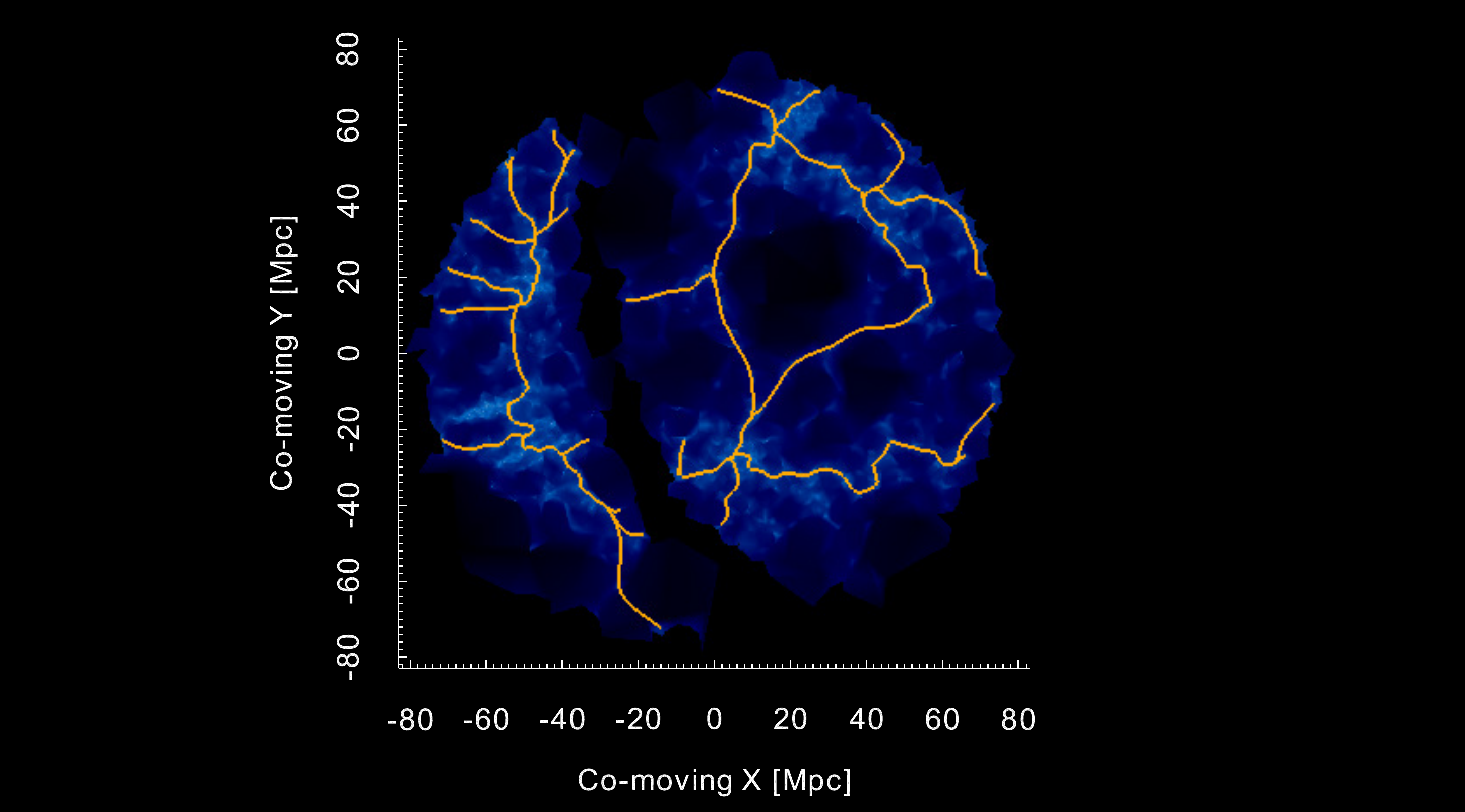}
\includegraphics[scale=.44, trim={5cm 0cm 7cm 0cm}, clip]{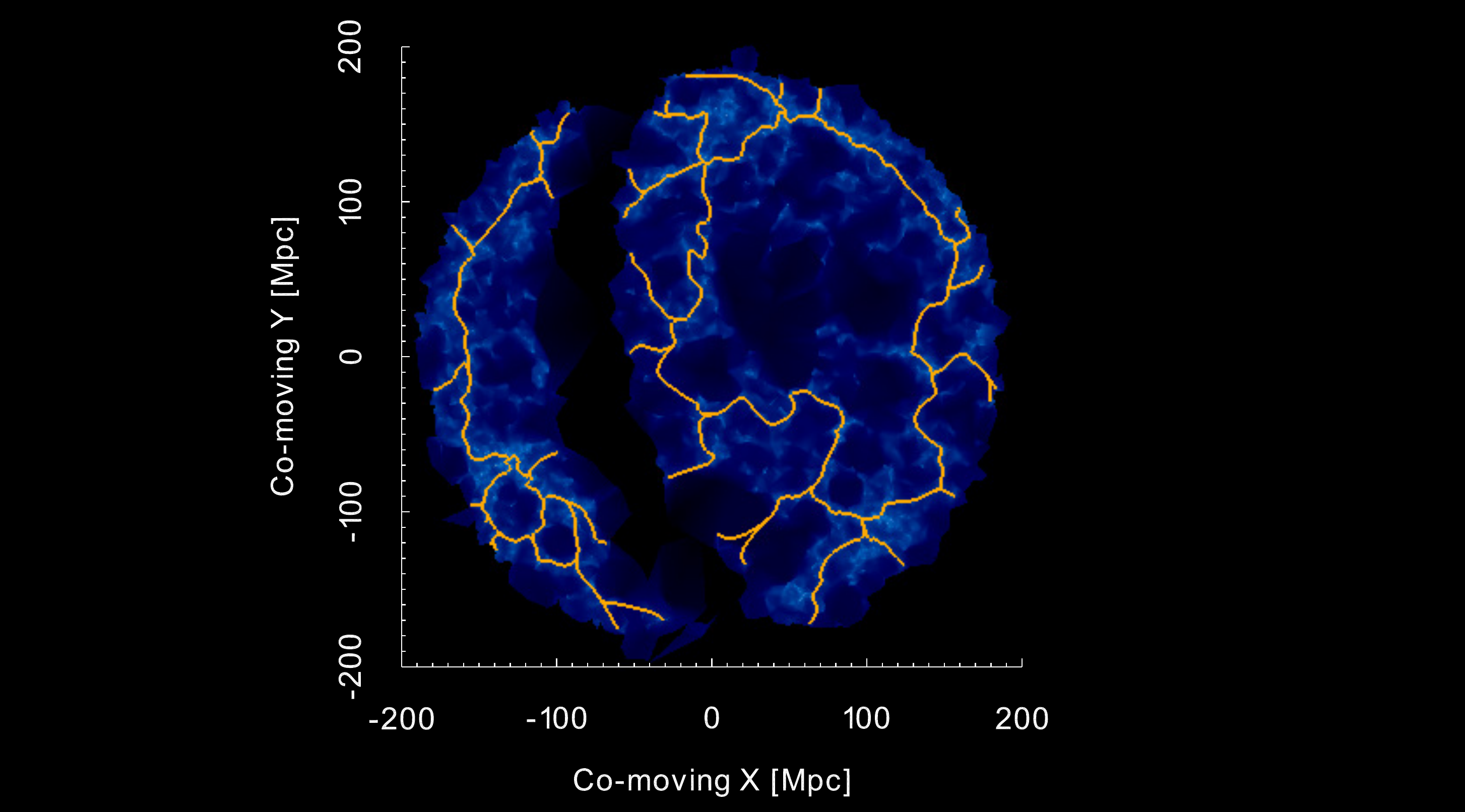}

\caption{The main steps of DisPerSE delineating filament backbones from a galaxy distribution. The left-hand column is the redshift slice 0.0125 $\leq$ $z$ $<$ 0.0175 and the right-hand column is the redshift slice 0.0375 $\leq$ z $<$ 0.0425. The top panel shows all 6dFGS galaxies for the respective redshift ranges and the middle panel shows the corresponding Delaunay tessellations. The tessellation is coloured according to the local density field where underdense regions are darker colours and overdensities are lighter colours. The bottom panel has the filament backbones (orange lines) found by DisPerSE overlaid on the tessellation. There is excellent agreement between the density of the tessellation and the galaxy distribution and excellent agreement between the thin overdensities in the tessellation and the filament backbones found by DisPerSE. This process for all the redshift slices can be viewed in Appendix \ref{appen:disperse}.}
\label{fig:disperse}
\end{figure*}

While DisPerSE is scale and parameter-free, the user is required to select a threshold (persistence ratio -- $\sigma_{\rm pr}$) to extract the significant and robust topological features above the sampling noise. The persistence ratio is a signal-to-noise measurement of how sensitive a topological feature (such as a filament) is to change. Features with high persistence ratios have greater topological significance above the sampling noise. Choosing a persistence ratio that eliminates spurious detections while retaining only the most prominent features of the cosmic web is the aim. The persistence ratio will differ according to the distribution and density of the data. To find the optimal persistence ratio for each slice, we inspected the filament backbones returned by DisPerSE using initial persistence ratios between 2$\sigma_{\rm pr}$ and 5$\sigma_{\rm pr}$. Table \ref{tab:pratios} shows the redshift ranges for each 6dFGS slice and the persistence ratio chosen for each side of the Galactic plane. Our requirement to separate local and large-scale environmental effects leads us to choose conservative persistence ratios. This will return the most \textit{topologically robust} filament backbones, by which we mean as those unambiguously above the noise.

\begin{table}
\centering
\caption{The best persistence ratio ($\sigma_{\rm pr}$) for the galaxy distributions above ($b$ $>$ 10$^{\circ}$) and below ($b$ $<$ $-$10$^{\circ}$) the Galactic plane in each 6dFGS redshift slice. The persistence ratios were chosen to return the most topologically robust filament backbones.}
\label{tab:pratios}
\begin{tabular}{c c c}
  Redshift Range & $\sigma_{\rm pr}$ for $b$ $>$ 10$^{\circ}$ & $\sigma_{\rm pr}$ for $b$ $<$ $-$10$^{\circ}$  \\
  \hline
 0.0025 $\leq z <$ 0.0075 & 3.8  & 4.0  \\
 0.0075 $\leq z <$ 0.0125 & 3.2  & 3.8  \\
 0.0125 $\leq z <$ 0.0175 & 4.0  & 4.4  \\
 0.0175 $\leq z <$ 0.0225 & 3.6  & 4.2  \\
 0.0225 $\leq z <$ 0.0275 & 4.0  & 3.8  \\
 0.0275 $\leq z <$ 0.0325 & 3.6  & 4.6  \\
 0.0325 $\leq z <$ 0.0375 & 4.0  & 4.4  \\
 0.0375 $\leq z <$ 0.0425 & 4.0  & 4.8  \\
\end{tabular}
\end{table}

The main caveat of DisPerSE is that Morse theory requires filaments to circumscribe voids. This can sometimes lead to a filament being defined in a region that is underdense and most likely incomplete. By choosing a persistence ratio that returns the most topologically significant filament backbones, we minimize the occurrence of filament backbones being defined in unrealistic underdense regions. There are $\sim$5\,000 galaxies in the samples used in this work (Section \ref{samples}). Of the order of $\sim$10 galaxies from incomplete regions are assigned to these samples and do not affect our results.

\subsection{Sample Selection}
\label{samples}

We create two samples:
\begin{enumerate}
\item The \textit{near-filament} sample containing 5\,358 galaxies with available stellar mass estimates, limited to 0.7 Mpc from filament backbones with a projected density of $\Sigma_{5}$ $<$ 3 galaxies Mpc$^{-2}$.

\item A \textit{control sample} containing 8\,149 galaxies with available stellar mass estimates, a minimum distance of 5 Mpc from the nearest filament backbone and projected density of $\Sigma_{5}$ $<$ 3 galaxies Mpc$^{-2}$.
\end{enumerate}

\begin{figure*}
    \includegraphics[scale = 1]{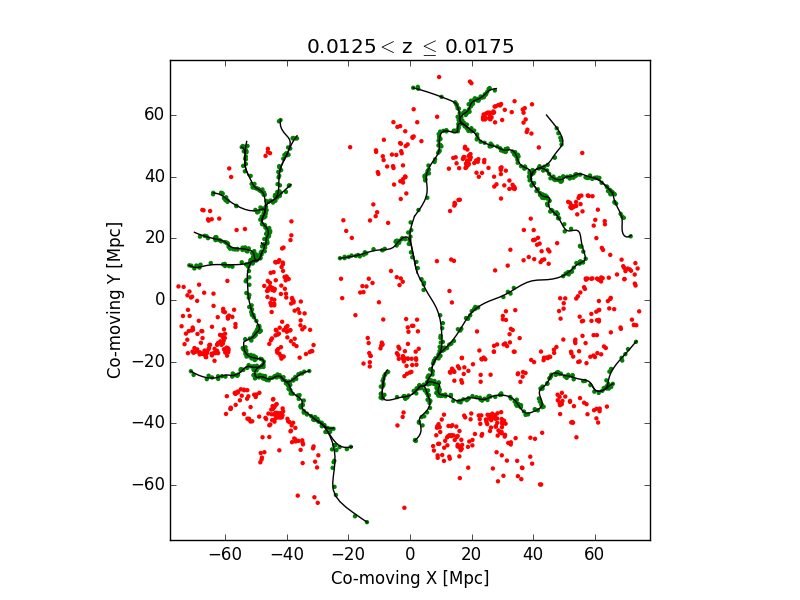}
    \caption{Redshift slice 0.0125 $\leq$ $z$ $<$ 0.0175 showing filament backbones (solid black lines), galaxies near filaments (green points) and the control sample (red points).}
    \label{fig:samp_eg}
\end{figure*}

The purpose of these two samples is to differentiate galaxies on the basis of surrounding large-scale structure and an example is shown in Fig.~\ref{fig:samp_eg}. A maximum distance of 0.7 Mpc was chosen for the near-filament sample as this is considered a thin radius for filaments of galaxies \citep{Tempel2014}. The minimum distance of 5 Mpc from the nearest filament backbone in the control sample implements a clear buffer between the two samples that eliminates potential in-falling galaxies in the control sample. There are 36\,805 6dFGS galaxies in the HIPASS volume where 8\,436 and 12\,119 are in the near-filament and control samples prior to any density or stellar mass restriction.

In each redshift slice we obtain a measure of local galaxy density using the fifth nearest neighbour projected density \citep[e.g.][]{Balogh2004, Brough2013}, defined as

\begin{equation}
	 \Sigma_{5}  = \frac{5}{\rm{\pi} r_{5}^{2}} ,
\label{eqn:proj_dens}
\end{equation}

where $r_{5}$ is the projected distance in Mpc to the fifth nearest galaxy from the target galaxy. We restrict both samples to have a maximum projected density of 3 galaxies Mpc$^{-2}$. This density threshold reduces the near-filament and control samples to 7\,510 and 11\,982 galaxies and removes galaxies residing in overdense environments that are subjected to cluster mechanisms.

\begin{figure}
    \includegraphics[width=\columnwidth]{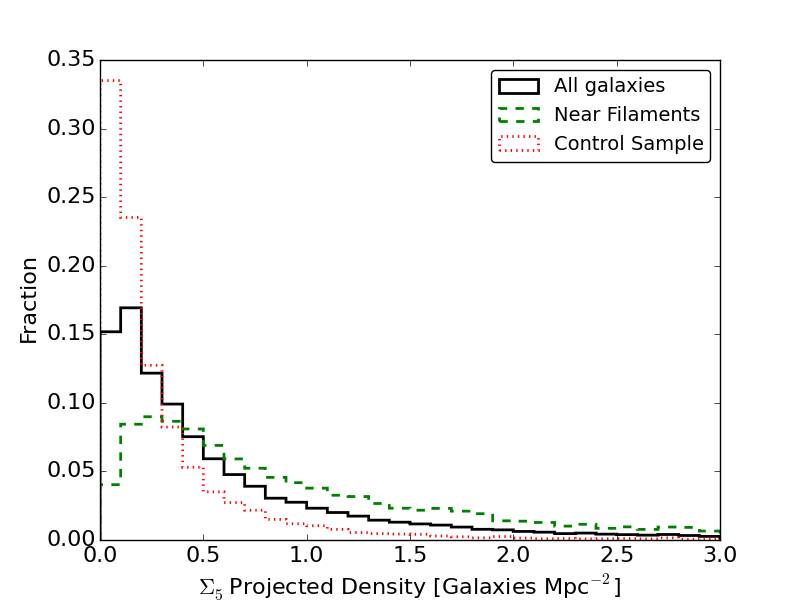}
    \caption{Normalized distribution of projected density for all 6dFGS galaxies in the HIPASS volume (solid black), galaxies near-filament (dashed green) and the control sample (dotted red). The near-filament sample systematically contains higher density local environments than the control sample, that is dominated by the lowest density local environments.}
    \label{fig:samp_dens}
\end{figure}

Fig.~\ref{fig:samp_dens} shows the distribution of $\Sigma_{5}$ for all galaxies within the HIPASS volume, the near-filament sample and the control sample. It is clear that the near-filament sample has a different $\Sigma_{5}$ distribution compared to the control sample. The distribution of $\Sigma_{5}$ for the near filament systematically consists of denser local environments which reflects that filaments can be viewed as intermediately dense environments that mainly host groups of galaxies. The distribution of  $\Sigma_{5}$ for the control sample is dominated by the lowest density local environments, which shows that we have obtained a good representation of field galaxies far from filament backbones. Table \ref{tab:samp_dens} highlights the differences between the distributions of $\Sigma_{5}$ for the two samples used in this work.

\begin{table}
\centering
\caption{Projected density ($\Sigma_{5}$), in units of galaxies Mpc$^{-2}$ for all galaxies in the HIPASS volume, the near-filament and control sample. All samples have been restricted to $\Sigma_{5}$ < 3 galaxies Mpc$^{-2}$ and the projected density distribution of the near-filament sample is distinctly different from the control sample.}
\label{tab:samp_dens}
\begin{tabular}{c c c c c}
  Sample & Ngal & Mean & Median & Standard deviation  \\
  \hline
All galaxies & 36\,805 & 0.57 & 0.36 & 0.59 \\
Near filament & 7\,510 & 0.90 & 0.58 & 0.71 \\
Control & 11\,982 & 0.30 & 0.16 & 0.39 \\
\end{tabular}
\end{table}

\begin{figure}
    \includegraphics[width=\columnwidth]{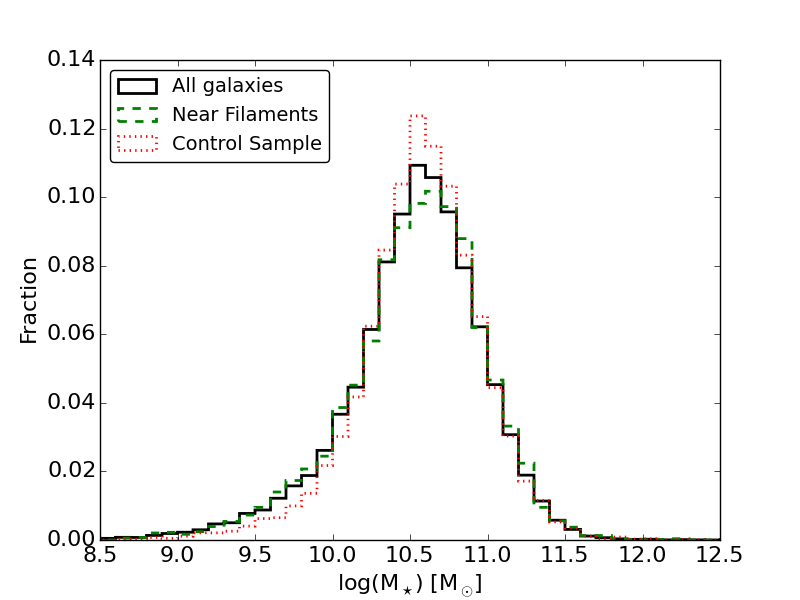}
    \caption{Normalized stellar mass distributions for all 6dFGS galaxies in the HIPASS volume (solid black), galaxies near filaments (dashed green) and the control sample (dotted red). The near-filament sample and control sample have similar stellar mass distributions.}
    \label{fig:samp_SM}
\end{figure}

Stellar masses are estimated using the $J$--band photometry \citep{Beutler2013} and the photometry of a galaxy is dependent on its input catalogue. As 6dFGS utilized multiple input catalogues, not all galaxies have the $J$--band photometry and these galaxies have no stellar mass estimate. This reduces the number of galaxies in the near-filament and control samples and our final samples contain 5\,358 and 8\,149 galaxies. Fig.~\ref{fig:samp_SM} shows the distribution of the stellar masses for all 6dFGS galaxies within the HIPASS volume, the near-filament and control samples. The stellar mass distributions of the near-filament and control samples show remarkable similarities. Table \ref{tab:samp_sm} quantifies the properties of the distributions.

\begin{table}
\centering
\caption{Stellar mass distributions in units of log(M${_\odot}$) for all 6dFGS galaxies in the HIPASS volume, the final near-filament and control samples. The near-filament and control sample distributions have very similar properties to each other.}
\label{tab:samp_sm}
\begin{tabular}{c c c c c}
  Sample & Ngal & Mean & Median & Standard deviation  \\
  \hline
All galaxies & 26\,855 & 10.52 & 10.56 & 0.45 \\
Near filament & 5\,358 & 10.53 & 10.58 & 0.46 \\
Control & 8\,149 & 10.57 & 10.58 & 0.39 \\
\end{tabular}
\end{table}

The indistinguishable stellar mass distributions and maximum projected density threshold of the near-filament and control samples ensure that we are comparing galaxies of the same stellar mass in similar local environments. 

\section{\HI Spectral stacking}
\label{stacking}
We perform spectral stacking in the same manner as \citet{Delhaize2013} with two differences; (i) we stack in velocity space (described below), and (ii) we exclude certain galaxies that meet our criteria for unreliable spectra (Section \ref{source_conf}). 

We use 6dFGS positions and velocities to extract the flux spectra from the HIPASS cubes for each galaxy to be stacked. In each \HI flux spectrum, a fourth-order polynomial (excluding the emission region) is subtracted from the baselines to account for noise due to standing waves internally reflected off telescope structures, or data reduction artefacts. All flux spectra are converted to a \HI mass per unit velocity through 

\begin{equation}
	\frac{M_{\rm H{\textsc i}}}{\Delta v}  = \frac{2.356 \times 10^{5}}{(1 + z)} \frac{S}{\Delta v} D_{\rm L}^{2} ,
\label{eqn:HI_mass}
\end{equation}

where $\frac{S}{\Delta v}$ is the flux per unit velocity (in Jansky km s$^{-1}$) and $D_{\rm L}$ is the luminosity distance (in Mpc). We need to convert spectra into the \HI domain prior to co-adding to use the \HI luminosity rather than flux. Each \HI spectrum is shifted to its respective rest frame that aligns all spectra to the same point of reference. The co-added weighted \HI mass per unit velocity of the sample is found as

\begin{equation}
	\left\langle \frac{M_{\rm H{\textsc i}}}{\Delta v} \right\rangle = \frac{\sum_{i=1}^{N} (w_{i} \frac{M_{\rm H{\textsc i}, \textit{i}}}{\Delta v})}{\sum_{i=1}^{N} w_{i}} ,
\label{eqn:stack}
\end{equation}

where $\frac{M_{\rm H{\textsc i}, \textit{i}}}{\Delta v}$ is the \HI mass per unit velocity of the $i^{\rm th}$ galaxy and $w_{i}$ is the associated weight. We calculate the weight of the $i^{\rm th}$ galaxy as

\begin{equation}
	w_{i} = (\sigma_{i} D_{\rm{L},i}^{2})^{-2} ,
\label{eqn:weight}
\end{equation}

where $\sigma_{i}$ is the rms of the flux spectrum excluding the emission region of the galaxy and $D_{\rm{L}, i}$ is the luminosity distance. This choice in weight ensures spectra with a high rms and non-detections of distant galaxies do not overpower any \HI emission in the stack. The average \HI mass of a sample of galaxies (equation \ref{eqn:HI_final}) is measured by integrating over the velocity interval where emission is present [$v_{1}$, $v_{2}$] in the co-added weighted \HI mass per unit velocity $\left\langle \frac{M_{H{\textsc i}}}{\Delta v} \right\rangle$,

\begin{equation}
	 M_{\rm H{\textsc i}}  = \int_{v_{1}}^{v_{2}} \left\langle \frac{M_{\rm H{\textsc i}}}{\Delta v} \right\rangle \ {\rm d}v .
\label{eqn:HI_final}
\end{equation}

The corresponding \HI fraction ($f_{\rm H{\textsc i}}$) is 

\begin{equation}
	 f_{\rm H{\textsc i}}  = \log_{10} \left( \frac{M_{\rm H{\textsc i}}}{M_{\rm{\star}}} \right) , 
\label{eqn:HI_frac}
\end{equation}

where $M_{\rm H{\textsc i}}$ is the \HI mass (in solar masses) measured from the stacking process and $M_{\star}$ is average stellar mass (in solar masses) of the sample. By measuring the \HI fraction, we are able to compare the \HI content of galaxies for different stellar masses. 

The dominant uncertainty in the \HI fraction is the uncertainty in the average \HI mass. To measure the uncertainty in the average \HI mass for a sample of $N$ galaxies, we stack $N$ `random' flux spectra. Each random flux spectrum is extracted using a random sky position and velocity (2\,000 $\leq$ $v_{\rm random}$ $\leq$ 11\,500 km s$^{-1}$), chosen to avoid Galactic emission and the edges of each HIPASS cube. Random flux spectra are stacked in the same way as actual co-added weighted \HI data.

The velocity range ($v_{1}$ and $v_{2}$) used in equation (\ref{eqn:HI_final}) is set by the boundaries of the stacked profile that cross the rms of the random stack for the first time. The rms of the randoms is integrated between $v_{1}$ and $v_{2}$ to measure the 1$\sigma$ uncertainty in the stack. 

If the average \HI mass 3 standard deviations above the uncertainty, it is considered a clear detection. Otherwise, there is no clear detection and the 1$\sigma$ upper limit is measured by integrating the rms of the randoms $\pm$ 150 km s$^{-1}$, a typical rotation velocity for galaxies in a \HI stacked sample  \citep{Delhaize2013}.

\subsection{Accounting for spectra with \HI flux of indeterminable origin}
\label{source_conf}
The full width half-maximum of the gridded HIPASS beam is $\sim$15.5 arcmin that can have multiple galaxies contributing \HI flux to a single spectrum \citep[e.g.~see][]{Doyle2005, Rohde2006}. The spectra with \HI flux of indeterminable origin above the HIPASS noise are considered \textit{unreliable}, which we exclude by constructing a simple mock \HI profile and determining if it would be observed in HIPASS.

Taking the HIPASS beam to be Gaussian, a rectangular mock \HI profile is constructed for any 6dFGS galaxy (except the central galaxy) within 3 standard deviations (19.8 arcmin) from the beam centre, a recessional velocity within $\pm$ 300 km s$^{-1}$ of the central galaxy and has a ($b_{\rm j}$ $-$ $r_{\rm F}$) colour $<$ 1.3. We have chosen the velocity overlap to be 300 km s$^{-1}$ because 600 km s$^{-1}$ is the maximum profile width of the HICAT galaxies \citep{Doyle2005} and in the simulations of \citet{Obreschkow2013}. ($b_{\rm j}$ $-$ $r_{\rm F}$) = 1.3 is the reddest colour of 6dFGS galaxies with HIPASS detections and we choose the same value as our upper limit to include H\,{\sc i}-rich galaxies likely to contribute flux in HIPASS spectra.

The width of the mock profile is predicted using the Tully--Fisher relation measured from HIPASS galaxies with known optical counterparts \citep[c.f.][]{Delhaize2013}. The $B$--band scaling relation of \citet{Denes2014} used to predict the \HI mass of a galaxy has the least scatter in the literature, which we use to estimate the integrated \HI mass for the mock \HI profile. The width of the \HI profile and the integrated \HI mass of the galaxy (area of the profile) enables us to determine the peak flux (amplitude of the profile) for the mock \HI profile. We convolve each mock \HI profile with a Gaussian weight in accordance with its angular distance from the assumed Gaussian beam centre.

\begin{figure}
    \includegraphics[width=\columnwidth]{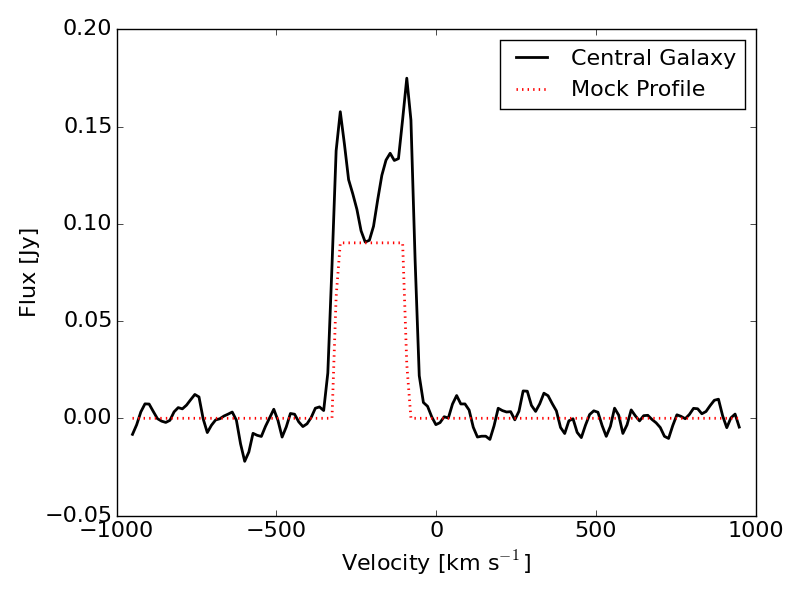}
    \caption{An example of an unreliable spectrum (black solid line) excluded from the stack. There is no emission from the central galaxy centred at 0 km s$^{-1}$. However, there is clear \HI emission from a nearby galaxy offset from the rest velocity of the central galaxy. The mock \HI profile (red dotted line) constructed to estimate the flux contribution of nearby galaxies is clearly above the noise and less than the peak flux of the spectrum. This spectrum is deemed unreliable and is subsequently removed from the sample.}
    \label{fig:conf}
\end{figure}

A spectrum is considered unreliable and excluded from the stack if the peak flux of the mock \HI profile is greater than 3 standard deviations of the rms and less than the peak flux of its HIPASS spectrum. Fig.~\ref{fig:conf} shows an example of a unreliable spectrum that has been excluded from the stack, in this way.

\begin{figure}
    \includegraphics[width=\columnwidth]{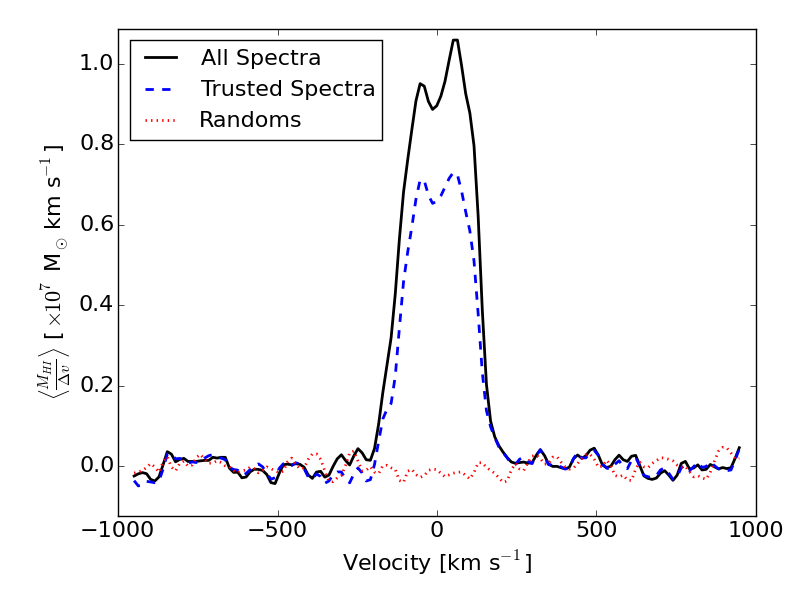}
    \caption{The greatest difference between stacking all 619 spectra (solid black line) and removing eight unreliable spectra (dashed blue line). The dotted solid red line is the random stack used to measure the uncertainty of the average \HI mass. The flux of the reliable stacked profile (dashed blue) has significantly decreased by $\sim$20\% compared to the stacked profile of all spectra (solid black). This decrease of flux is consistent with the expected contribution of neighbouring H\,{\sc i}-rich galaxies.}
    \label{fig:conf_stack}
\end{figure}

Fig.~\ref{fig:conf_stack} shows the stacked spectrum with the greatest difference between including all spectra and excluding unreliable spectra. Eight out of a total of 619 spectra were removed from the stack and resulted in an $\sim$20\% decrease of \HI flux. A decrease of \HI flux is expected as neighbouring H\,{\sc i}-rich galaxies contribute additional flux to the stack. While unreliable spectra can make a significant difference (Fig.~\ref{fig:conf_stack}), it does not dominate the \HI flux of the stack. The majority of 6dFGS galaxies are not detected above the HIPASS noise as unreliable spectra typically make up a few percent of any sample and the flux from neighbouring galaxies offset from the beam centre quickly decrease due to the Gaussian beam response.

\section{Results: \HI fraction for varying stellar mass and projected densities}
\label{results}
We present the \HI fraction (H\,{\sc i}-to-stellar mass ratio) of galaxies for varying stellar masses and projected densities in the near-filament and control samples. Both stellar mass and local environment are known to affect the \HI fraction of a galaxy \citep[e.g.][]{Baldry2008, Catinella2010, Cortese2011, Brown2015} and we separate these quantities in subsamples to determine if the large-scale filamentary environment affects the \HI fraction of galaxies with the same stellar mass and projected density.

\begin{figure*}
\centering
\includegraphics[scale=.46]{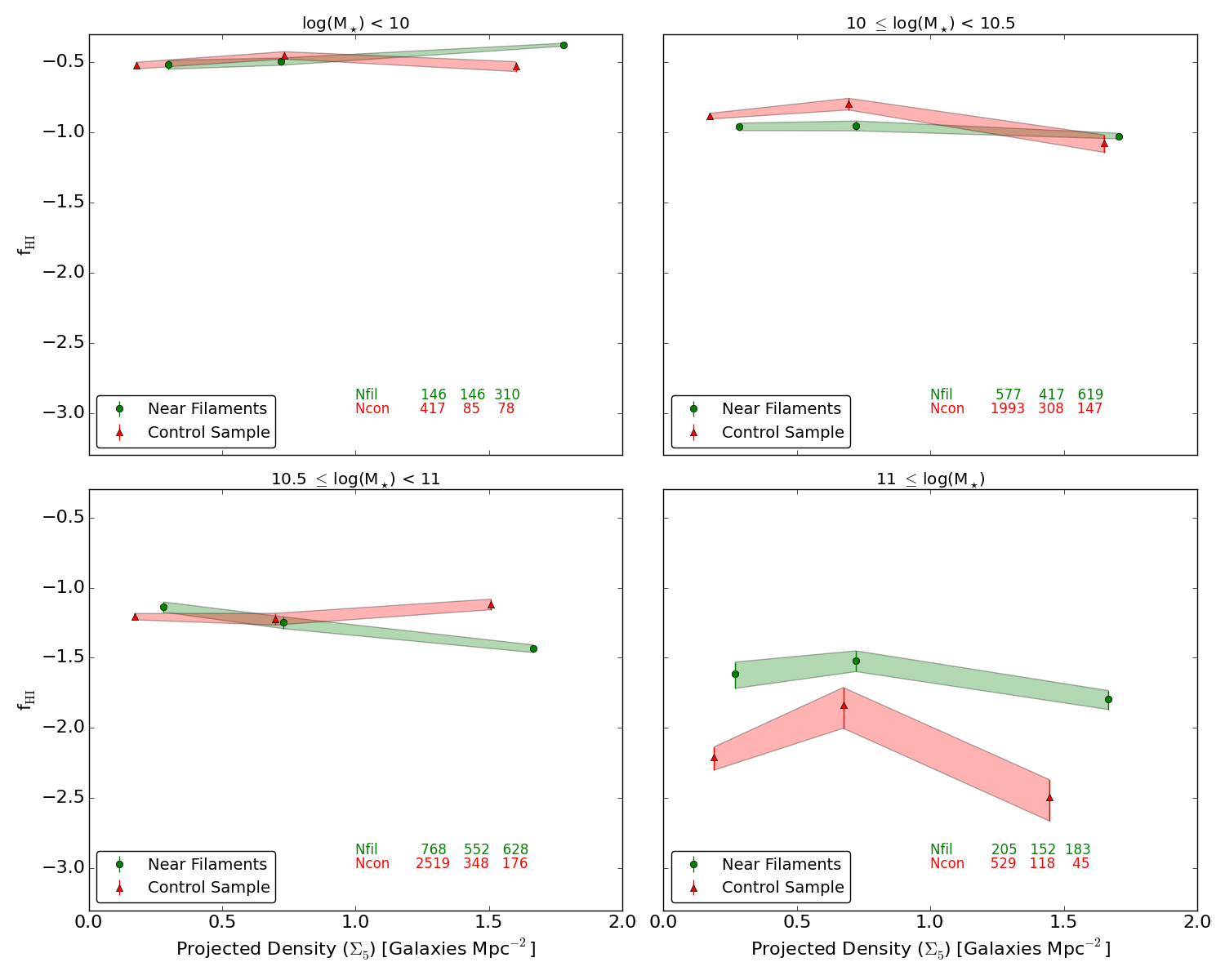}
\caption{The \HI fraction in the near-filament sample (green circles) and control sample (red triangles) for varying mean projected densities in four different stellar mass ranges. The number of galaxies in each stack are shown in text with the same colours and the shaded regions signify the 1$\sigma$ uncertainties derived from the rms and width of the stacked profile. The \HI fraction as a function of projected density decreases for increasing stellar mass ranges in both the near-filament and control samples. For stellar masses log($M_{\star}$) $<$ 11 M$_{\odot}$, the \HI fractions are flat and indistinguishable from each other at the 3$\sigma$ level. For stellar masses log($M_{\star}$) $\geq$ 11 M$_{\odot}$, the near-filament \HI fraction is significantly higher than the control sample by up to 0.75 dex with a 5.5$\sigma$ significance. This difference occurs at a mean projected density of $\Sigma_{5}$ = 1.45 galaxies Mpc$^{-2}$.}
\label{fig:HI_SM_per_dens}
\end{figure*}

Fig.~\ref{fig:HI_SM_per_dens} shows the \HI fraction in the near-filament and control samples for varying projected densities in four different stellar mass ranges. The \HI fractions decrease with increasing stellar mass range. For galaxies with stellar masses log($M_{\star}$) $<$ 11 M$_{\odot}$, the \HI fractions of the near-filament samples is within 0.1 dex of the control samples. In this mass range, the \HI fractions are flat at the 3$\sigma$ level. For galaxies with stellar masses log($M_{\star}$) $\geq$ 11 M$_{\odot}$, the \HI fraction of the near-filament sample is systematically greater than the control sample. It is statistically significant at the 4.9$\sigma$ and 5.5$\sigma$ level where the difference is 0.6 and 0.75 dex at mean projected densities of 0.2 and 1.45 galaxies Mpc$^{-2}$. Assuming the \HI followed this trend, the difference between the near-filament and control samples would be even more significant ($\sim$1 dex) at a mean projected density of 1.8 galaxies Mpc$^{-2}$.

\section{Discussion}
\label{discussion}
Two ideas currently being tested in galaxy evolution are: (i) whether galaxies near filaments evolve differently to galaxies far from filaments, (ii) if any difference in evolution be explained purely by stellar mass and local galaxy density.

In the CWD model, stellar mass and local galaxy density are primary drivers of detaching galaxies from the cosmic web, ending cold gas accretion and quenches star formation. Our results suggest that galaxies with log($M_{\star}$) $\geq$ 11 M$_{\odot}$ near filaments have not (or only recently) detached from the cosmic web and are still accreting cold gas from the cosmic web. Galaxies of the same mass in the same projected densities have already detached from the cosmic web and no longer have access to the cold gas of the cosmic web. This supports the natural motivation of the CWD model \citep{Aragon-Calvo2016} and could be evidence for cold mode accretions \citep{Keres2005} with massive galaxies replenishing their gas content through cold gas accretion from the cosmic web. 

This accretion does not enhance their gas content to the extent of making them gas rich, the low-mass galaxies are clearly the most gas rich (highest \HI fraction). Instead, the accretion gives them a slightly enhanced gas content compared to field galaxies of the same mass that are unable to accrete gas from filaments. If galaxies near filaments stay attached to the cosmic web for longer periods than galaxies far from filaments (as our results suggest), gas accretion from the cosmic web may be responsible for a burst of star formation observed on the outskirts of clusters \citep{Porter2008, Mahajan2013}. Star formation generally traces \HI and is primarily correlated with stellar mass \citep[e.g.][]{Doyle2006}. As cosmic web gas is diffuse in the local Universe, any accretion would be second order compared to the \HI and star formation driven by stellar mass and local galaxy density. Extra gas accreted from the cosmic web may require dense environments to trigger an appreciable difference in star formation.

We only observe a significant difference in \HI fractions in the most massive galaxies in our sample. These galaxies near filaments are located at the deepest minima of the large-scale gravitational potential which would funnel the intrafilament gas on to the galaxies. The less massive galaxies have smaller gravitational potentials and no evidence of cold gas accretion from the large-scale structure is observed.

\citet{Wang2015} show H\,{\sc i}-rich galaxies are surrounded by a H\,{\sc i}-rich environment where the gas is situated in clumps, hosted by small satellite galaxies or density peaks in the diffuse \HI intergalactic medium. These observations also support the CWD model and are best explained by cold mode accretion. The H\,{\sc i}-rich galaxies in \citet{Wang2015} are rare, and their contribution to the \HI fractions in this study is averaged out. We are unable to directly compare our results to \citet{Wang2015} as their galaxies did not exceed log($M_{\star}$) $=$ 11 M$_{\odot}$, where we see evidence for gas accretion from filaments. However, both this work and \citet{Wang2015} support the CWD model \citep{Aragon-Calvo2016} and cold mode accretion \citep{Keres2005}.  

The \HI content of galaxies can be influenced by local environment where the galaxy \HI mass decreases with increasing local density \citep[e.g.][]{Cortese2011, Serra2012, Catinella2013}. We do not observe a decreasing \HI fraction for increasing projected densities as we impose  maximum projected density of 3 galaxies Mpc$^{-2}$ for galaxies in both our samples. This eliminates galaxies in high-density environments (such as dense groups and clusters) where the aforementioned environmental trend is clearest. Therefore, we are probing low-to-intermediate projected densities and do not measure \HI fractions where strong environmental mechanisms remove \HI gas from galaxies. The \HI fraction as a function of projected density is flat and indistinguishable between the near-filament and control samples for stellar masses log($M_{\star}$) $<$ 11 M$_{\odot}$ (Fig.~\ref{fig:HI_SM_per_dens} top- and bottom-left panels). The only \HI fraction as a function of projected density that is different between the near-filament and control samples is for galaxies with stellar masses log($M_{\star}$) $\geq$ 11 M$_{\odot}$ (Fig.~\ref{fig:HI_SM_per_dens} bottom-right panel). As we have accounted for stellar mass and local environment, we attribute the difference in \HI fractions to a galaxies proximity to filaments. 

Previous studies have shown galaxies close to filaments possess the same properties \citep{Alpaslan2015} or very similar properties \citep{Chen2015b, Guo2015, Alpaslan2016, Martinez2016} to galaxies of the same stellar mass but not filament members. The main conclusion from these studies is that any differences due to their filamentary environment are second-order effects. While these studies were conducted in the optical and UV regime, we draw the same conclusion with the measured \HI fractions for the majority of our sample with only high-mass galaxies having a statistically different \HI fraction but is still second order in comparison to the correlation of stellar mass with \HI fraction.

The work of \citet{Chen2015a} show that the morphology--density relation can extend to the scale of filaments. As the galaxy density is higher in filaments than the field, it is reasonable to expect more (typically gas poor) early types in the near-filament sample than the control sample. However, filaments host individual and small galaxy groups where the morphology and gas content depend on the dynamical history. If a filament galaxy has spent sufficient time in a dense group, it may have a bulge-dominated morphology with no supply of cold gas. In this scenario, the galaxy has detached from the cosmic web as a result of strong  group interactions. In principle, this may occur more in filaments than the field. However, filaments are environments of intermediate density and groups of this density are rare. Hence, interactions of this magnitude would not dominate the galaxy population in filaments. Even though the projected density of the samples used in this work is significantly different (Fig.~\ref{fig:samp_dens}), neither sample is inherently gas rich (or poor) as a result of its local environment.

Other studies \citep[e.g.][]{Alpaslan2015, Martinez2016} find that the star formation rate and specific star formation rate is slightly lower in filament galaxies compared to the field. These trends are weaker when contrasting the star formation of cluster and field galaxies. Both filament and field galaxies span the full spectrum of gas poor to gas rich and passive to star forming. Similar to the gas content, star formation depends on the complex interplay of stellar mass, local galaxy density and cosmic web attachment. All these factors should be considered when discerning the reason behind a galaxies gas content and star formation. 

The \HI fractions measured in this work rely on HIPASS observations that detected only the brightest \HI galaxies in the nearby ($z <$ 0.043) Universe. It is predicted that WALLABY \citep{Koribalski2012} will observe \HI in 6 $\times$ 10$^{5}$ galaxies out to a redshift of $z <$ 0.26 and DINGO will detect up to 10$^{5}$ galaxies out to a redshift of $z <$ 0.43 \citep{Duffy2012}. WALLABY, DINGO and future SKA observations will be able to follow up \HI measurements of log($M_{\star}$) $=$ 11 M$_{\odot}$ galaxies near filaments that will provide better measurements of their \HI fraction to determine if this accretion scenario is plausible.

\section{Conclusions}
\label{conclusions}

Using 6dFGS and HIPASS, we compare the \HI fraction of galaxies near-filament backbones to galaxies in a control sample far from filaments. The two samples are created using perpendicular filament distance and a maximum projected density that excludes galaxies in high-density environments. The two samples have similar stellar mass distributions but significantly different projected density distributions. The \HI fraction is measured using spectral stacking and we vary the stellar mass and projected density for both samples to disentangle what influences the \HI gas in these galaxies. We find that:

\begin{itemize}
	\item \HI fraction is strongly correlated with a galaxies stellar mass, where low-mass galaxies have high \HI fractions compared to massive galaxies. While the rate at which galaxies build stellar mass depends on environment, the stellar mass can be used to estimate the \HI fraction to first order. Local and large-scale environments can also influence the \HI fraction but these are second-order effects compared to the stellar mass.

	\item We observe no significant difference in the \HI fraction for varying projected densities between the near-filament and control samples for galaxies with stellar masses log($M_{\star}$) $<$ 11 M$_{\odot}$. In this mass range, the \HI fractions for the near-filament and control sample are indistinguishable from each other in all stellar mass subsamples. This implies that filaments have no influence of the \HI content of galaxies compared to the correlation between \HI content, stellar mass and local galaxy density. 
		
	\item Galaxies near filaments with a stellar mass log($M_{\star}$) $\geq$ 11 M$_{\odot}$ have a systematically higher \HI fraction for varying projected densities than the same mass galaxies in the control sample. This difference is most pronounced at projected densities of $\Sigma_{5}$ = 0.2 and 1.45 galaxies Mpc$^{-2}$ where the statistical significance is 4.9$\sigma$ and 5.5$\sigma$. As stellar mass and local environment have been accounted for, we suggest that this is tentative evidence of massive galaxies accreting cold gas from the intrafilament medium. Galaxies in the control sample do not have access to this gas reservoir and would be unable to replenish their cold gas supply. Only the most massive galaxies in filaments have a large enough gravitational potential to draw gas from the intrafilament medium.
	
	\item We observe no trend (at the 3$\sigma$ level) in the \HI fractions for varying projected densities where the near-filament and control samples are indistinguishable from each other. Only galaxies with a projected density of 3 galaxies Mpc$^{-2}$ were included in the near-filament and control samples, which confines this study to low and intermediate local environments. This threshold is implemented specifically to exclude high-density (i.e.~cluster) environmental mechanisms where galaxies are typically \HI deficient. The maximum density threshold is the reason we do not see a decreasing \HI fraction with increasing projected density. 
	
\end{itemize}

Overall, the near-filament and control samples have the same \HI fraction as each other when varying the stellar mass and projected density. The exception arises for galaxies with stellar masses log($M_{\star}$) $\geq$ 11 M$_{\odot}$. Our results are best explained by the CWD model \citep{Aragon-Calvo2016} that allows galaxies in this mass range to utilize cold mode accretion \citep{Keres2005}. Through this mechanism, galaxies are able to accrete gas from the intrafilament medium and replenish their cold gas supply. This evidence is tentative and follow-up observations with more sensitive radio telescopes (such as ASKAP and SKA) will be required to statistically support the proposed scenario. 

This work is a pilot study for future redshift surveys and radio telescopes. Specifically, TAIPAN \citep{Hopkins2014} will reveal the cosmic web and low-mass galaxies in the southern sky in much greater detail than 6dFGS. ASKAP and SKA will change our understanding of the \HI gas in galaxies where our analysis can be directly applied to WALLABY \citep{Koribalski2012} and DINGO. We plan to use these surveys to follow up our work and improve our understanding of the \HI content of galaxies within the cosmic web. 

\section*{Acknowledgements}
This work is based on the third and final data release (DR3) of 6dFGS (\url{http://www.6dfgs.net/}).

The Parkes telescope is part of the Australia Telescope, which is funded by the Commonwealth of Australia for operation as a National Facility managed by CSIRO.

DK acknowledges support from an Australian post-graduate award (APA) and thanks the University of Hull for their hospitality where some of this work was undertaken. DK would like to thank Jacinta Delhaize and Toby Brown for insightful discussions regarding spectral stacking and \HI fractions. DK would also like to thank Theirre Sousbie for his help with the usage of DisPerSE. Finally, DK thanks the referee for improving the foundation and clarity of this work.

This project has received funding from the European Research Council (ERC) under the European Union's Horizon 2020 research and innovation programme (grant agreement no. 679627).

%%%%%%%%%%%%%%%%%%%%%%%%%%%%%%%%%%%%%%%%%%%%%%%%%%



\bibliographystyle{mnras}
\bibliography{References} % if your bibtex file is called example.bib

%%%%%%%%%%%%%%%%% APPENDICES %%%%%%%%%%%%%%%%%%%%%

\appendix

\section{Filament Backbones found with DisPerSE}
\label{appen:disperse}

Here, we show the filament backbones in eight 6dFGS redshift slices obtained with DisPerSE as described in Section \ref{fil_6dFGS}. The galaxies in each redshift slice are separated in two distinct sample -- above ($b$ $>$ 10$^{\circ}$) and below ($b$ $<$ $-$10$^{\circ}$) the Galactic plane, which ensures that DisPerSE does not spuriously connect filament backbones across the Galactic plane. A density field is produced through computing the Delaunay tessellation on each galaxy distribution. Discrete Morse theory is directly applied to the Delaunay tessellation, identifying topologically important features. Filament backbones are recovered as long, thin topologically important features and a persistence ratio (i.e.~signal-to-noise threshold, Table \ref{tab:pratios}) is chosen to retain filament backbones above the desired threshold. The persistence ratio chosen should reflect the filament backbones that can be visually verified in the galaxy distributions.

% Slice 0 and 1
\begin{figure*}
\centering
\includegraphics[scale=.45, trim={5cm 0cm 7cm 0cm}, clip]{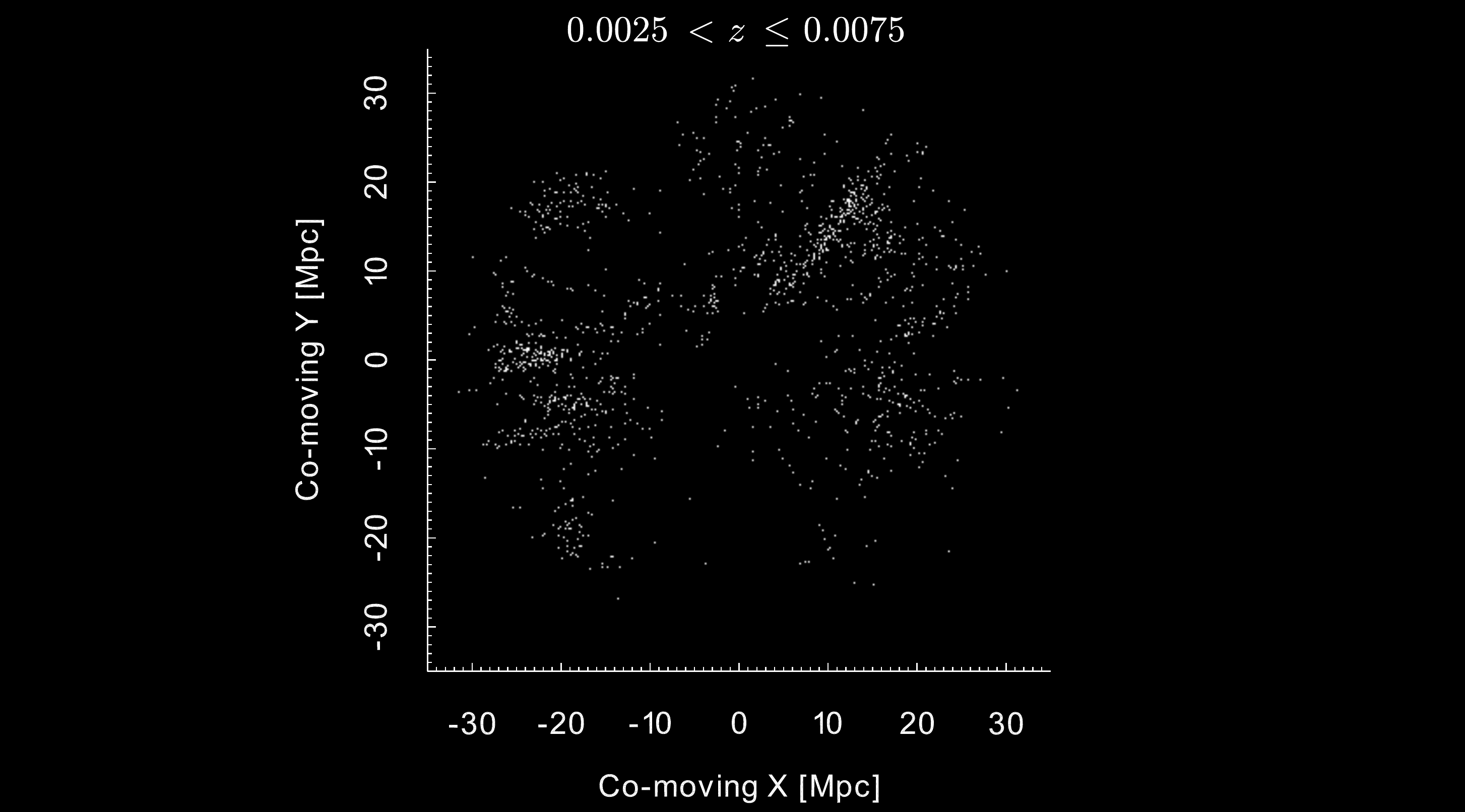}
\includegraphics[scale=.45, trim={5cm 0cm 7cm 0cm}, clip]{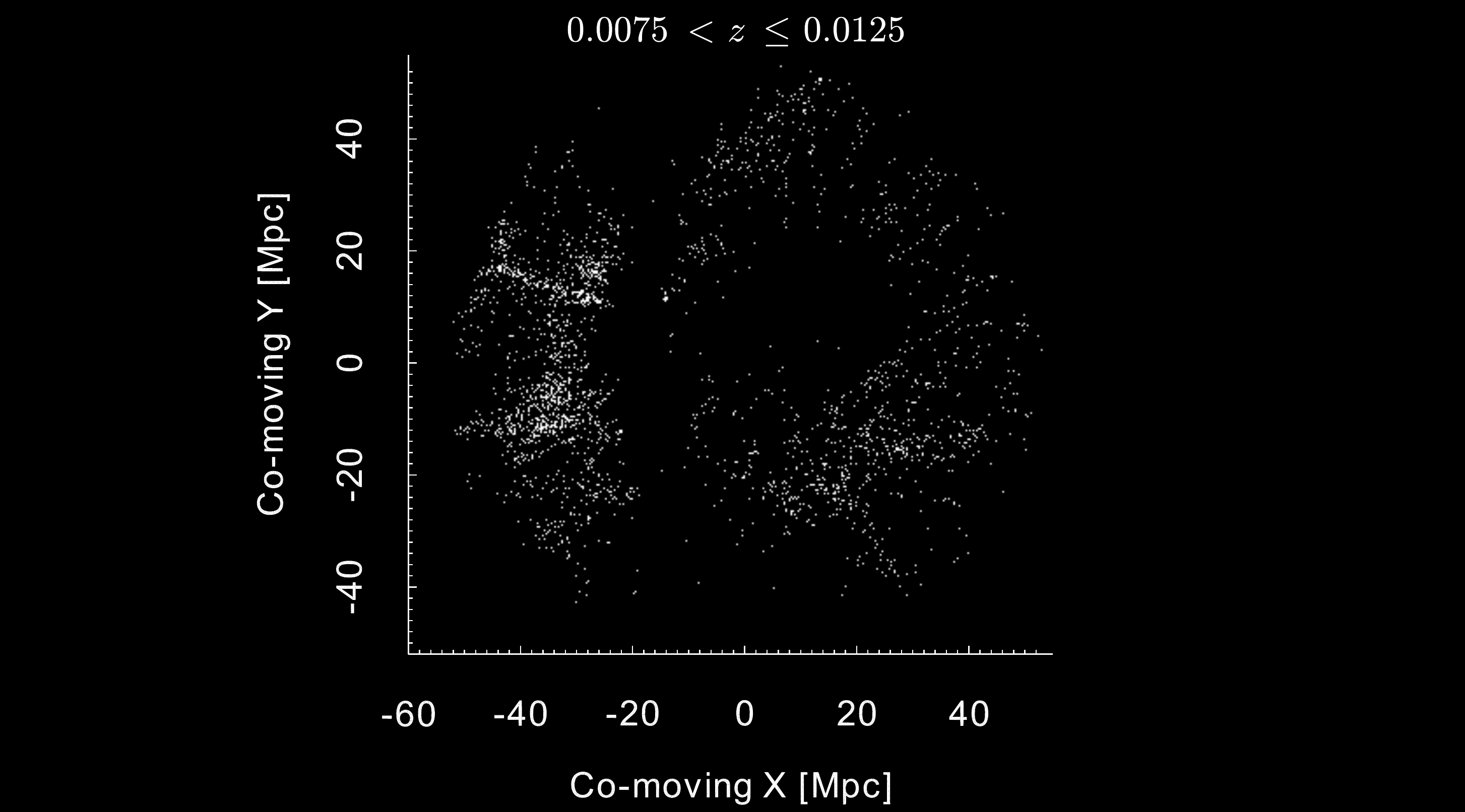}

\includegraphics[scale=.45, trim={5cm 0cm 7cm 0cm}, clip]{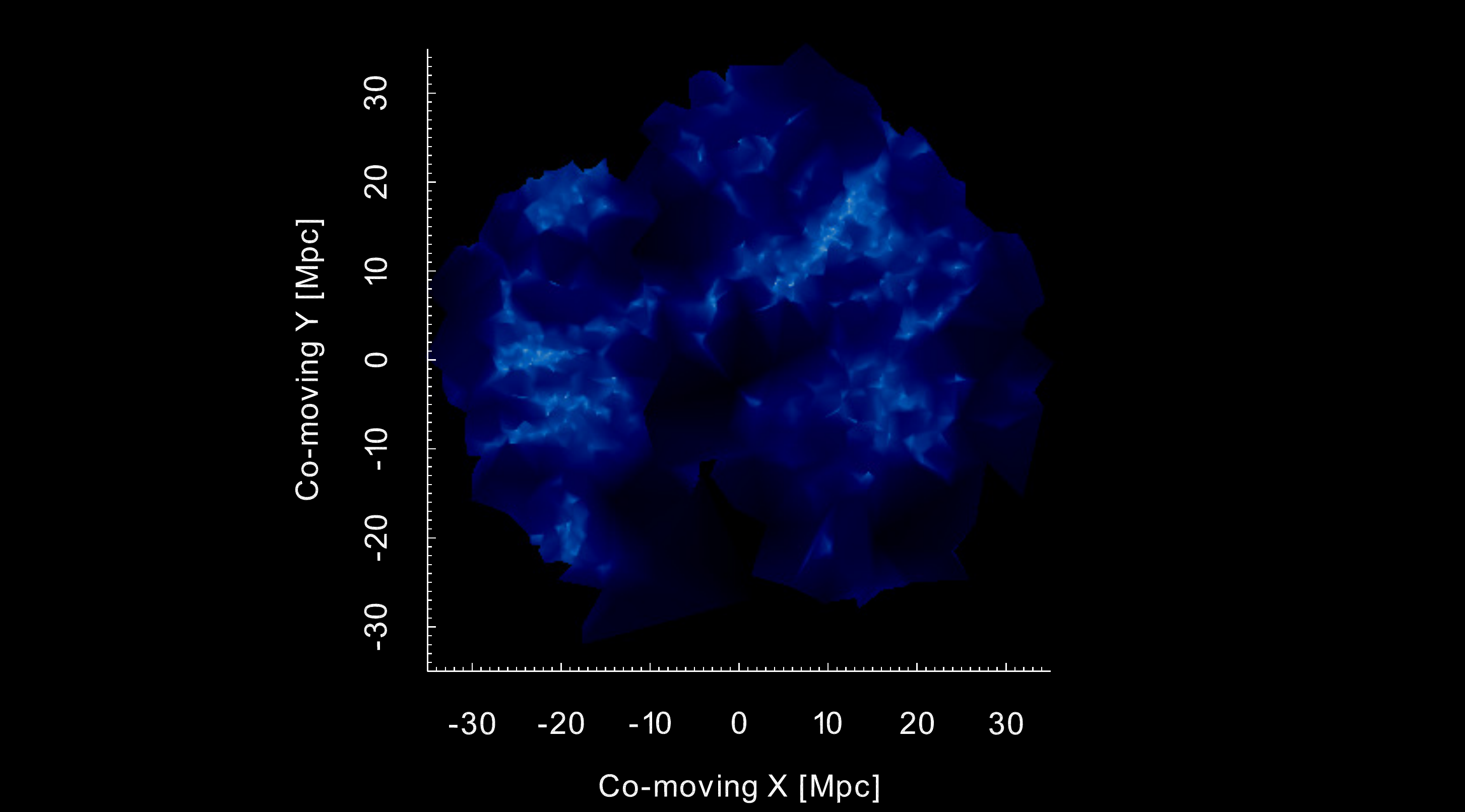}
\includegraphics[scale=.45, trim={5cm 0cm 7cm 0cm}, clip]{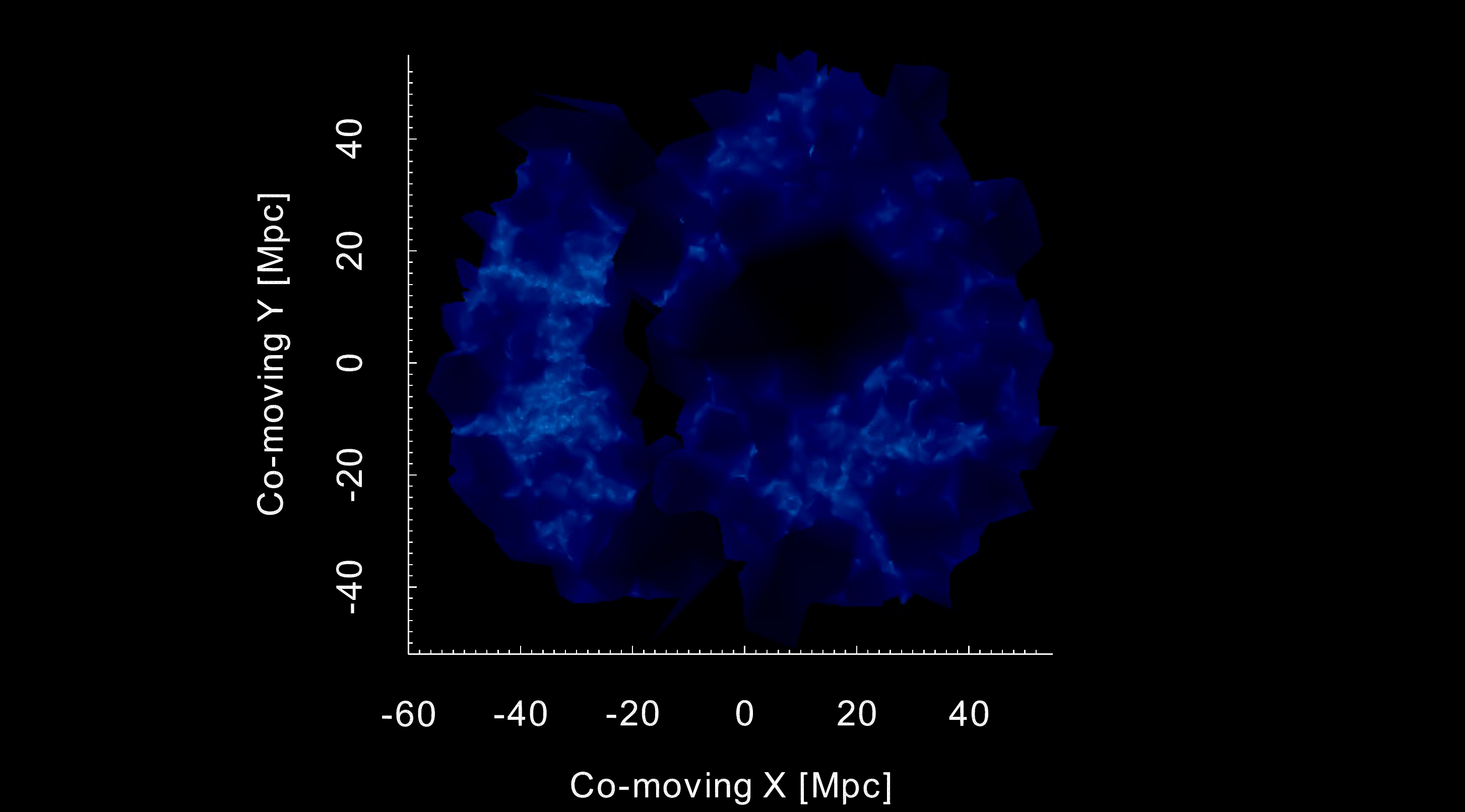}

\includegraphics[scale=.45, trim={5cm 0cm 7cm 0cm}, clip]{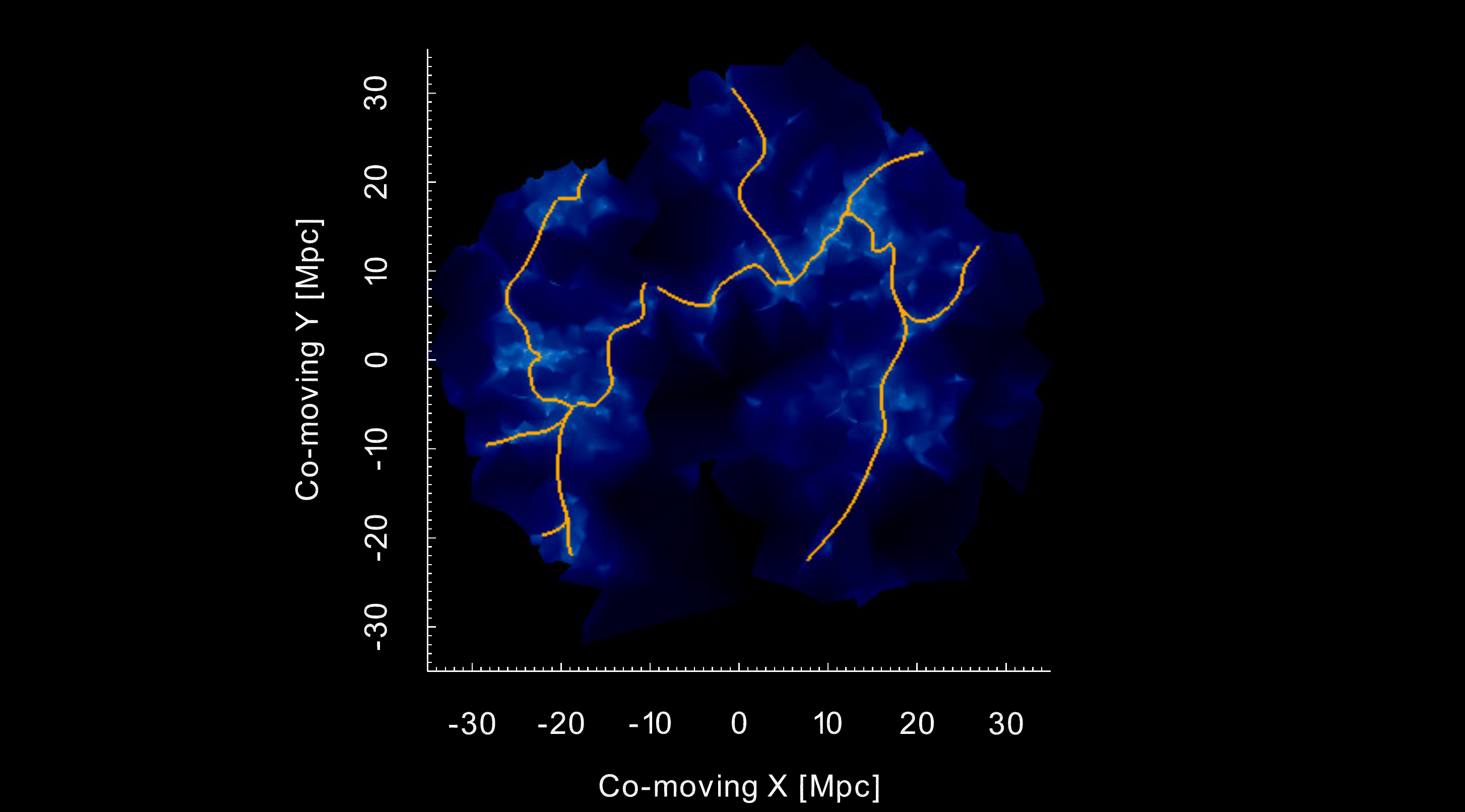}
\includegraphics[scale=.45, trim={5cm 0cm 7cm 0cm}, clip]{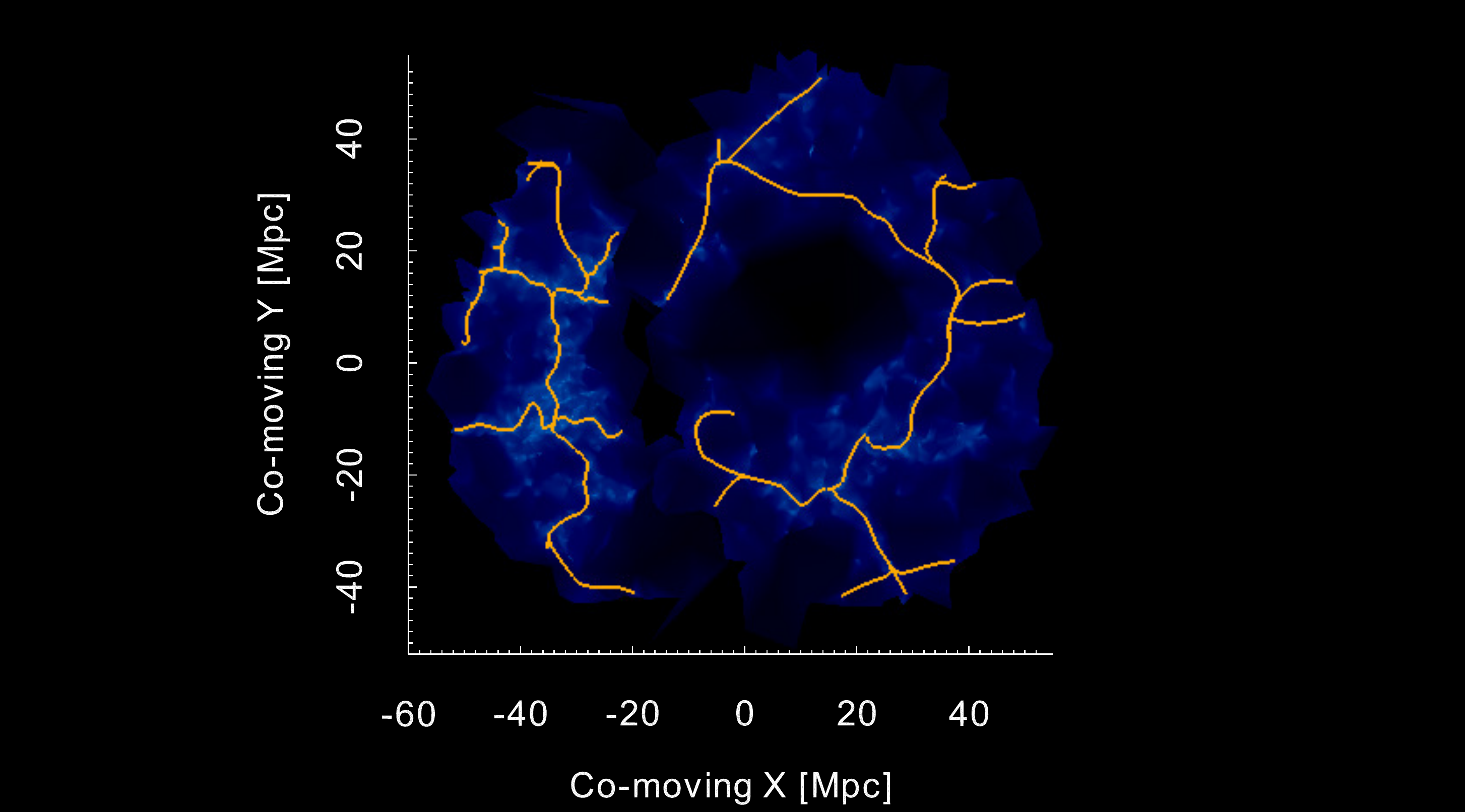}
\caption{Same as Fig.~\ref{fig:disperse} showing the main steps of DisPerSE delineating filament backbones in redshift slices 0.0025 $\leq z <$ 0.0075 (left-hand column) and 0.0075 $\leq z <$ 0.00125 (right-hand column). The top panel shows all 6dFGS galaxies for the respective redshift ranges and the middle panel shows the corresponding Delaunay tessellations. The tessellation is coloured according to the local density field where underdense regions are darker colours and overdensities are lighter colours. The bottom panel has the filament backbones (orange lines) found by DisPerSE overlaid on the tessellation.}
\label{fig:disperse01}
\end{figure*}

% Slice 2 and 3
\begin{figure*}
\centering
\includegraphics[scale=.45, trim={5cm 0cm 7cm 0cm}, clip]{Galaxies_s2b-eps-converted-to}
\includegraphics[scale=.45, trim={5cm 0cm 7cm 0cm}, clip]{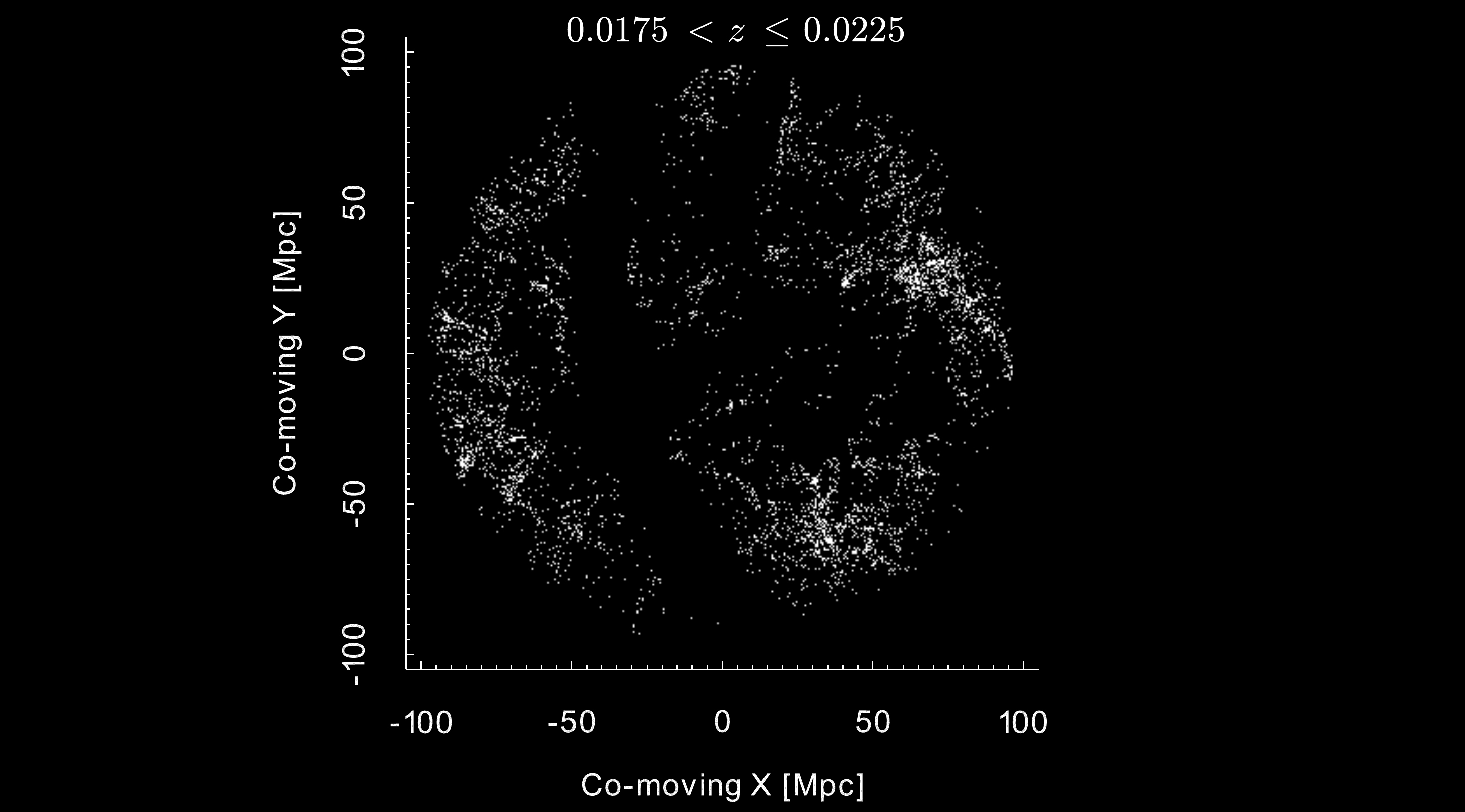}

\includegraphics[scale=.45, trim={5cm 0cm 7cm 0cm}, clip]{Delauany_tessellation_s2b-eps-converted-to}
\includegraphics[scale=.45, trim={5cm 0cm 7cm 0cm}, clip]{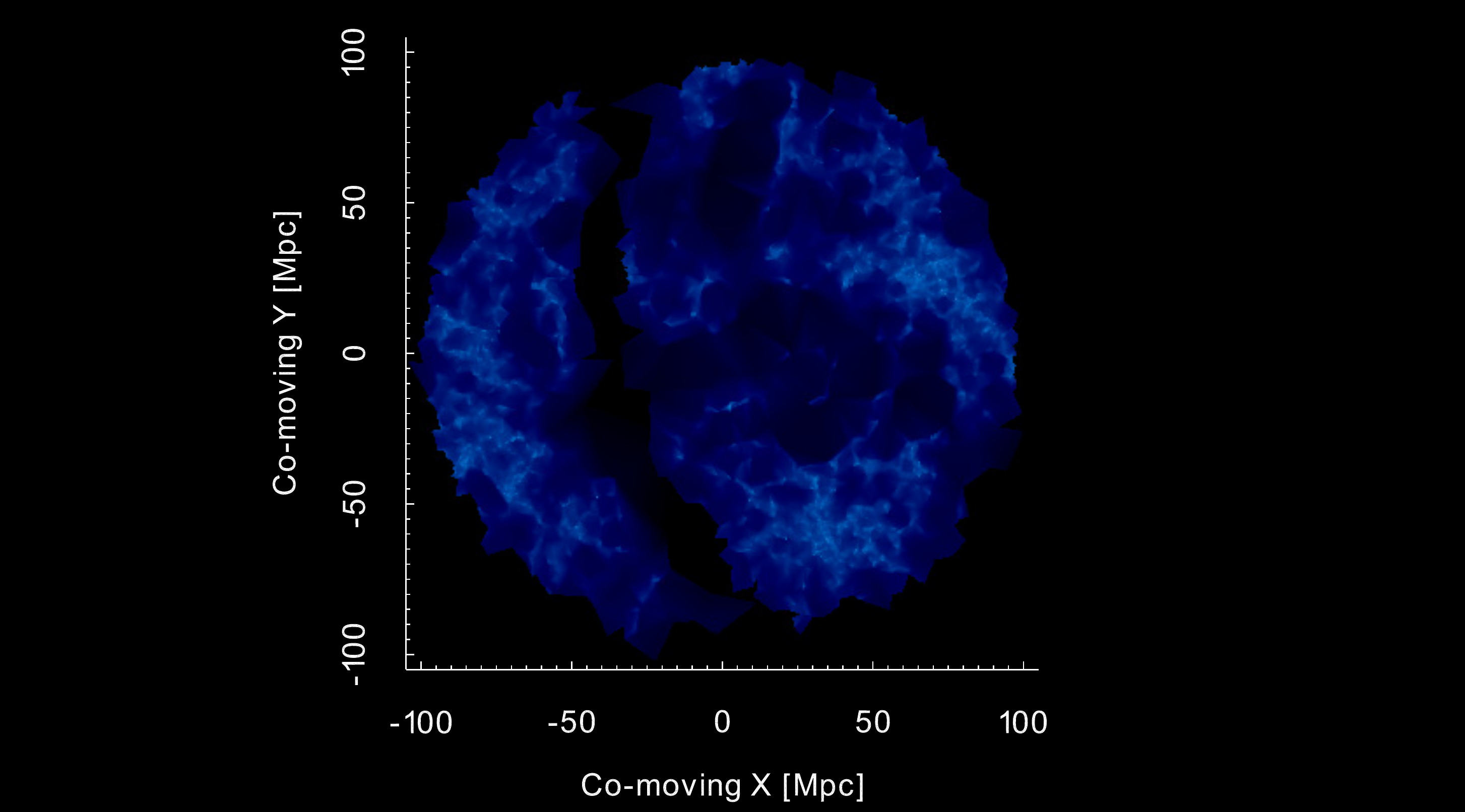}

\includegraphics[scale=.45, trim={5cm 0cm 7cm 0cm}, clip]{Filament_backbones_s2b-eps-converted-to}
\includegraphics[scale=.45, trim={5cm 0cm 7cm 0cm}, clip]{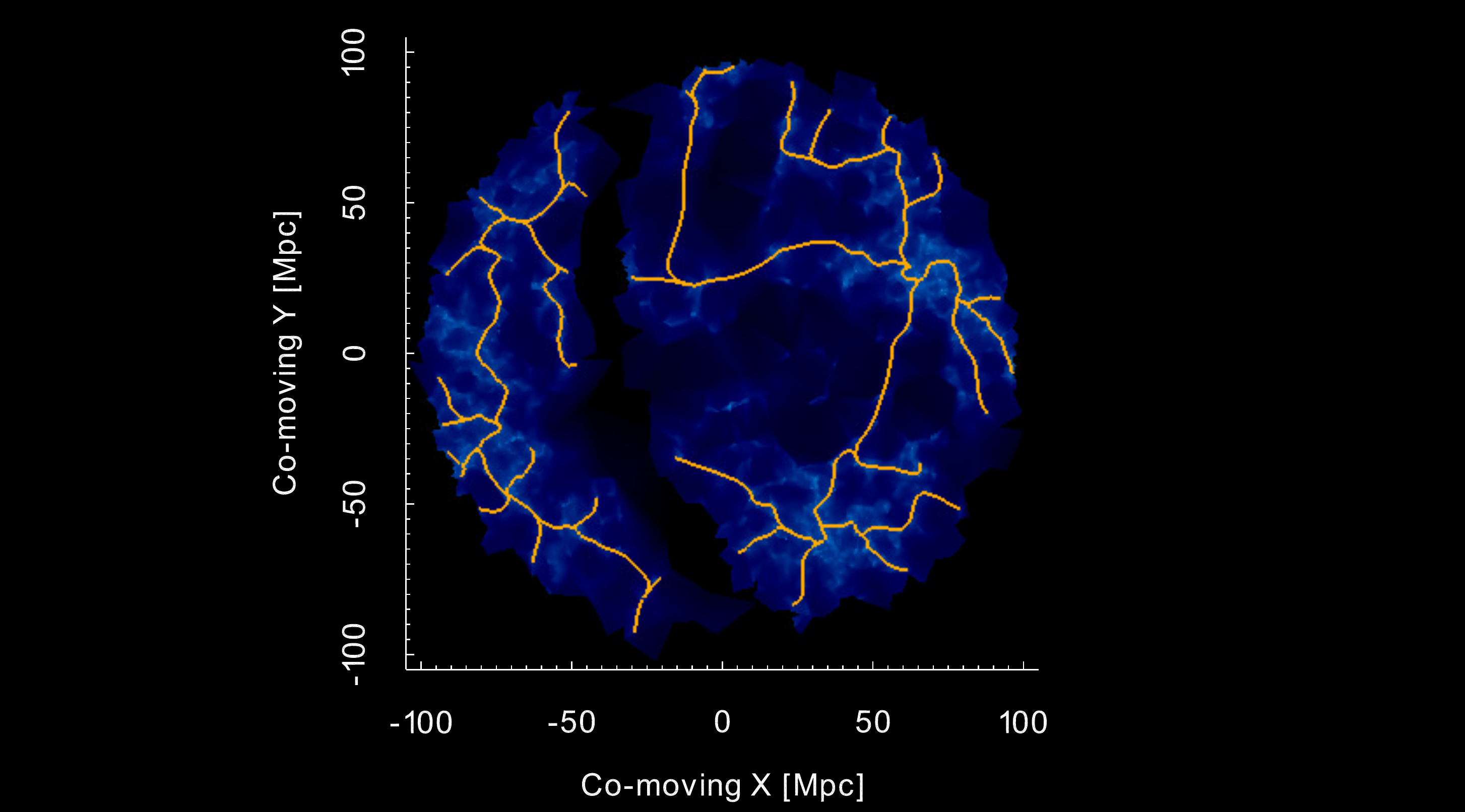}
\caption{Same as Fig.~\ref{fig:disperse01} for the redshift slices 0.0125 $\leq z <$ 0.0175 (left-hand column) and 0.0175 $\leq z <$ 0.00225 (right-hand column).}
\label{fig:disperse23}
\end{figure*}

% Slice 4 and 5
\begin{figure*}
\centering
\includegraphics[scale=.45, trim={5cm 0cm 7cm 0cm}, clip]{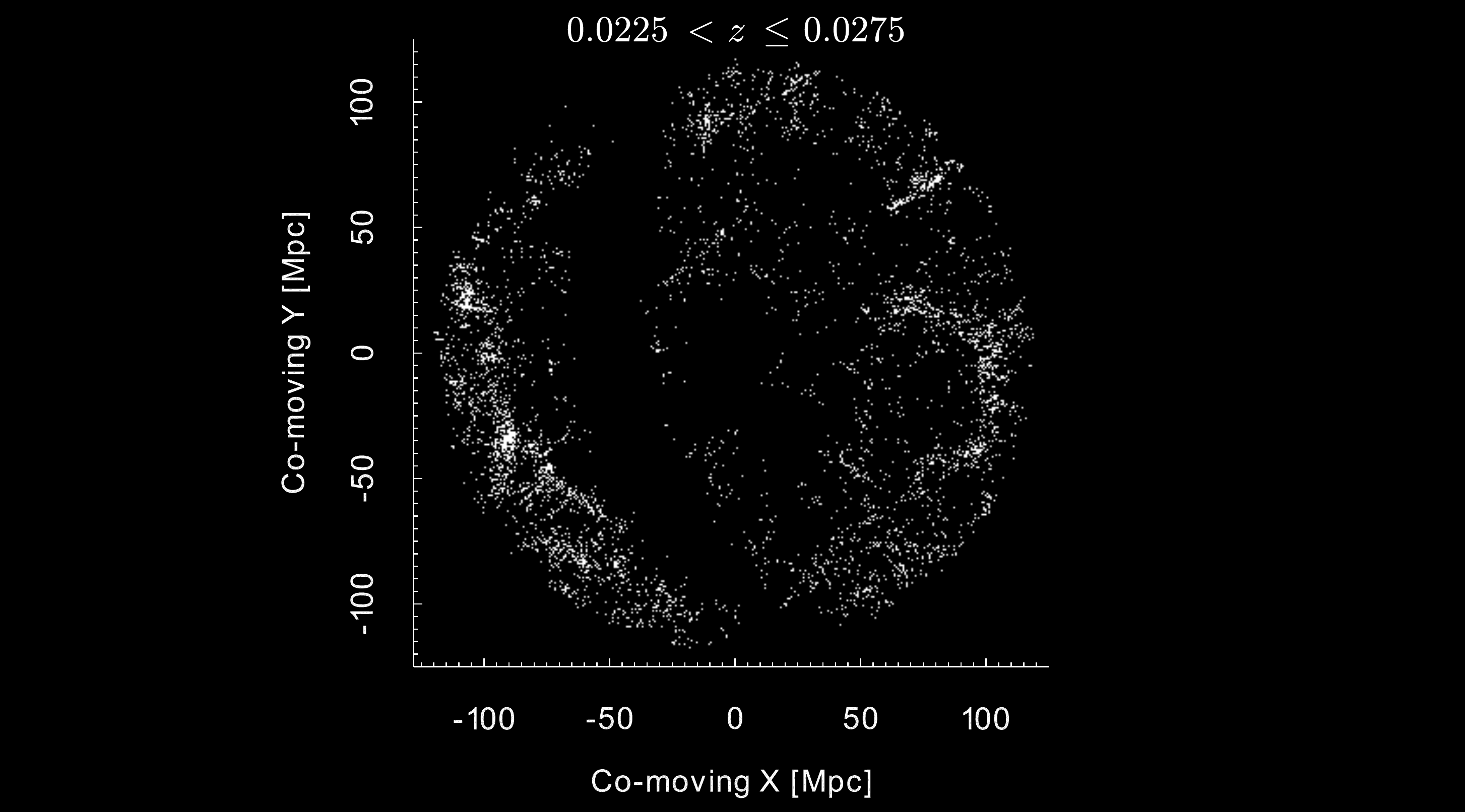}
\includegraphics[scale=.45, trim={5cm 0cm 7cm 0cm}, clip]{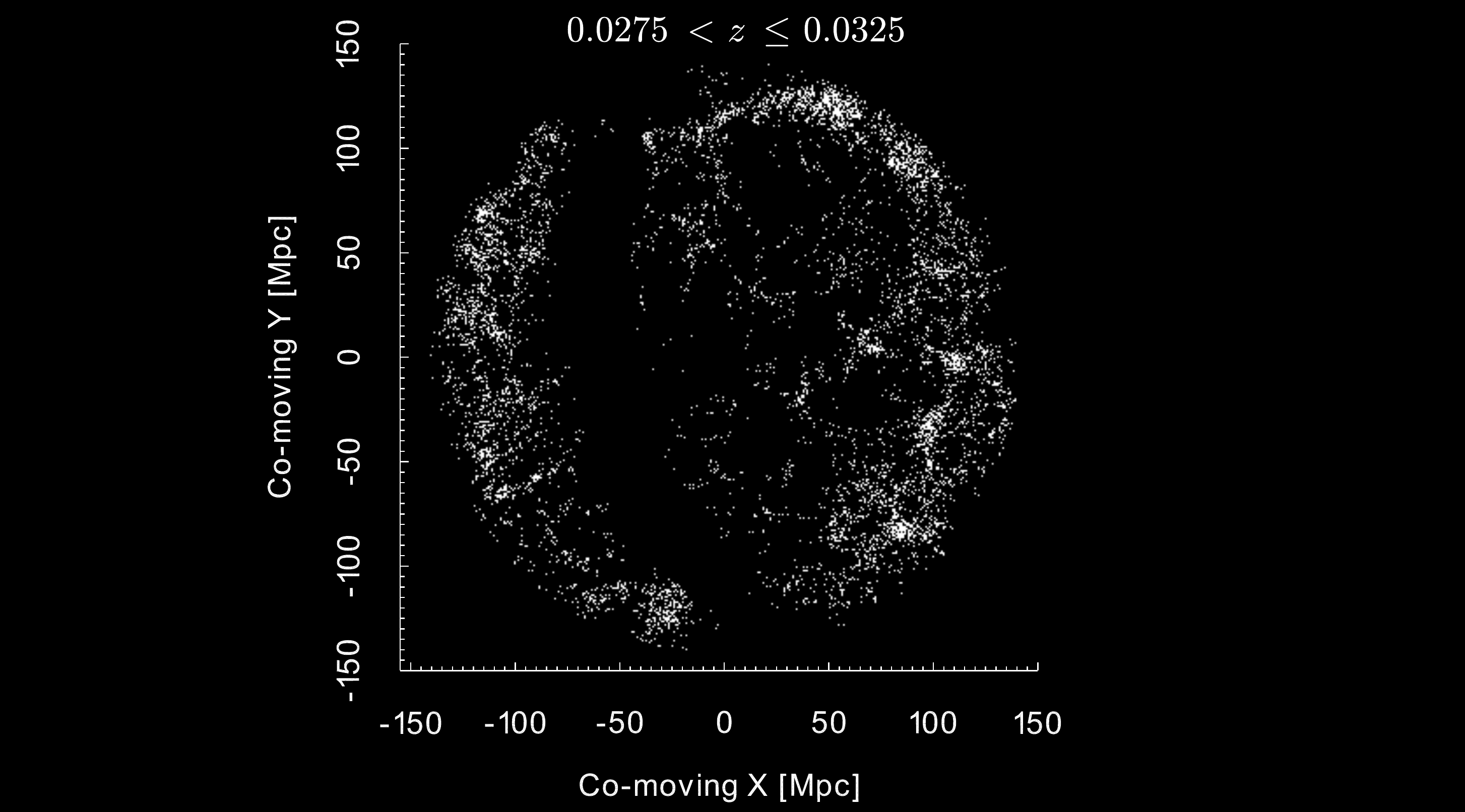}

\includegraphics[scale=.45, trim={5cm 0cm 7cm 0cm}, clip]{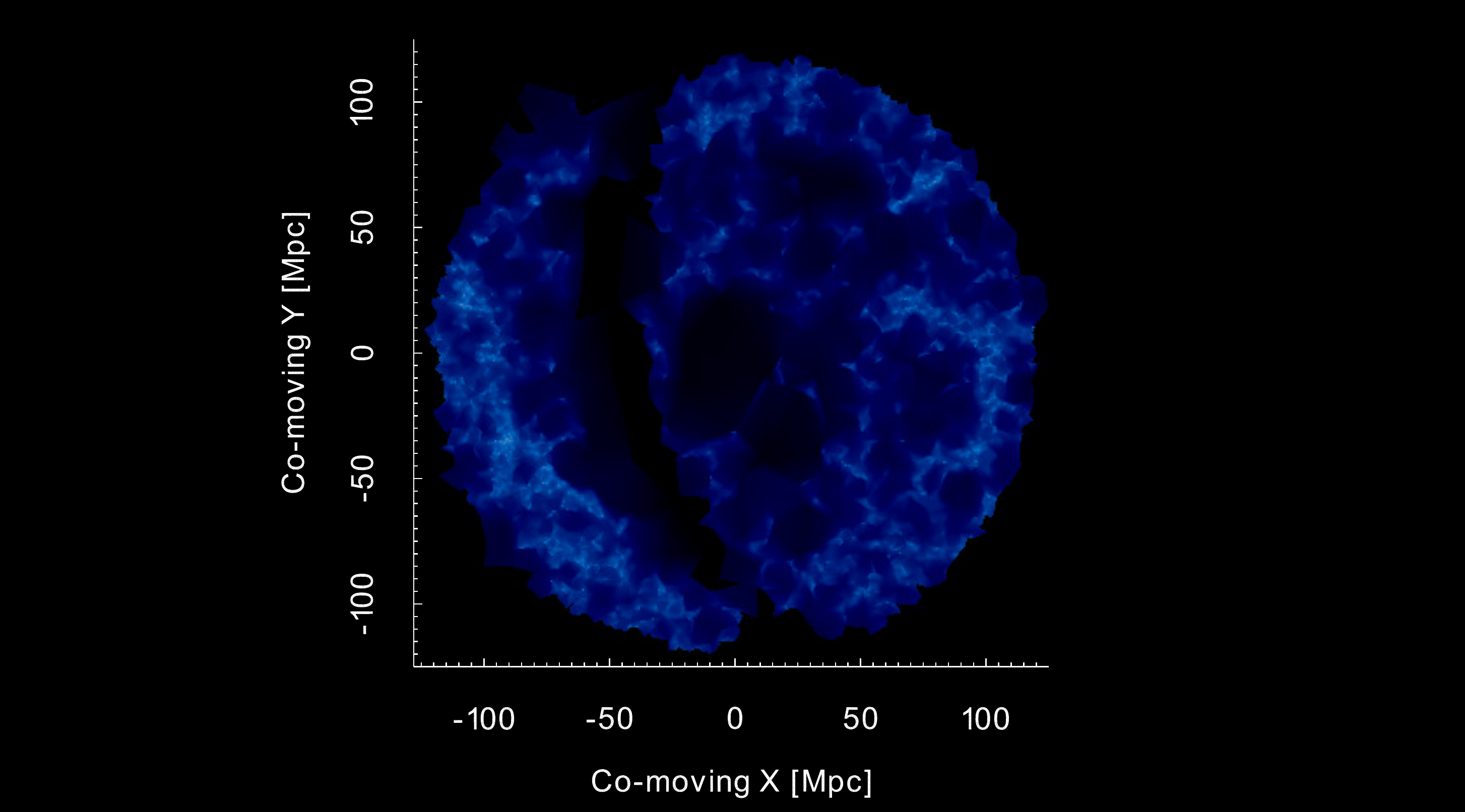}
\includegraphics[scale=.45, trim={5cm 0cm 7cm 0cm}, clip]{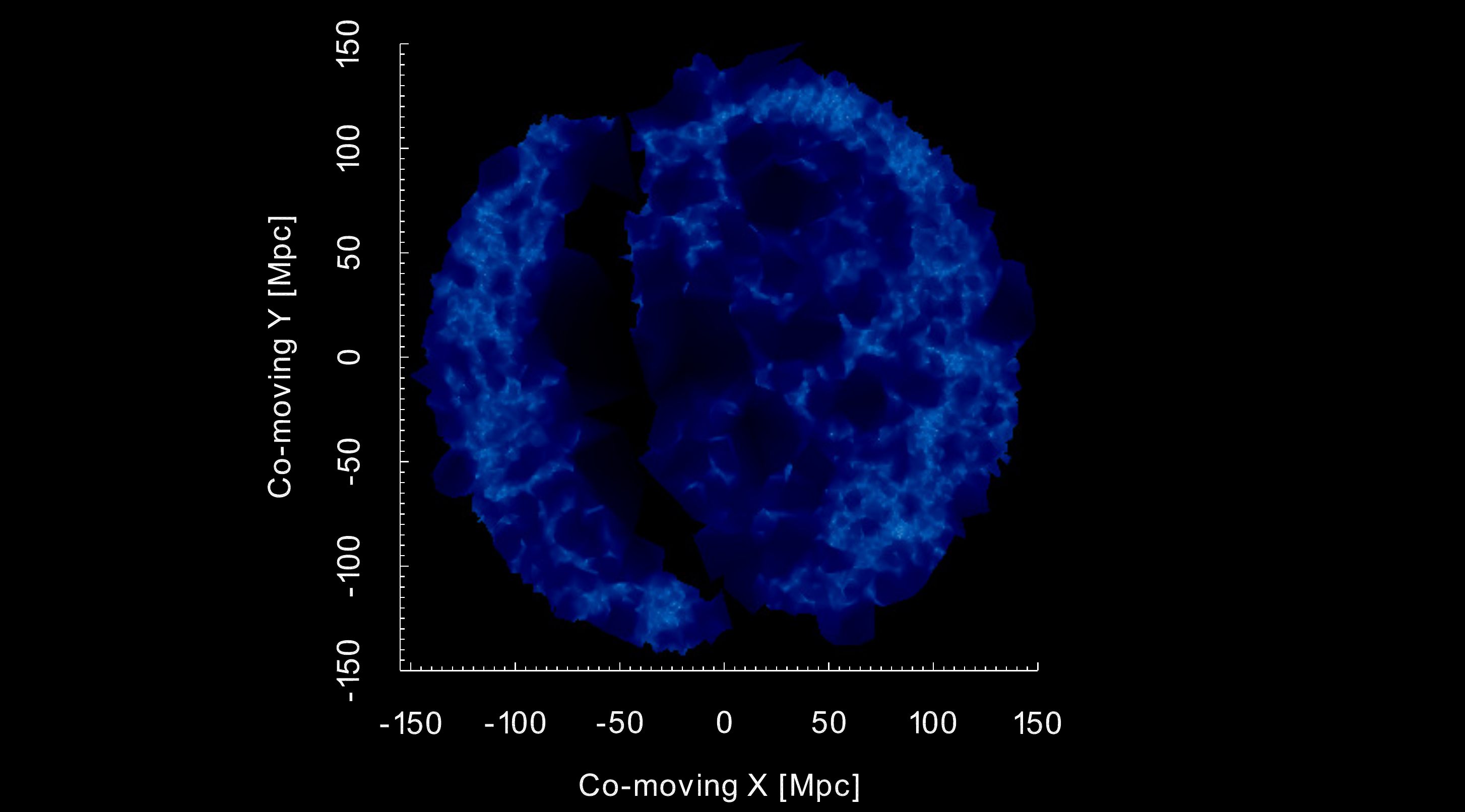}

\includegraphics[scale=.45, trim={5cm 0cm 7cm 0cm}, clip]{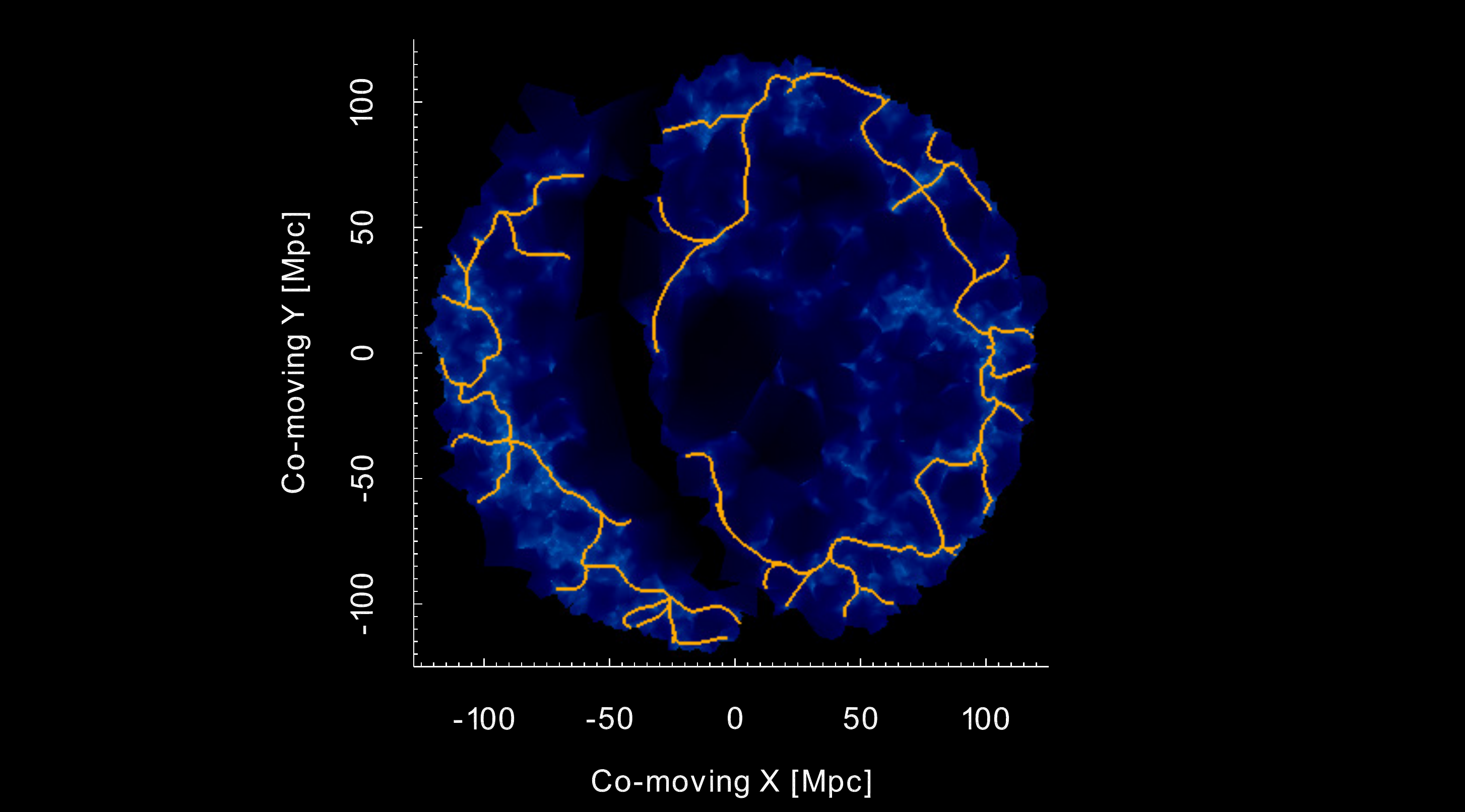}
\includegraphics[scale=.45, trim={5cm 0cm 7cm 0cm}, clip]{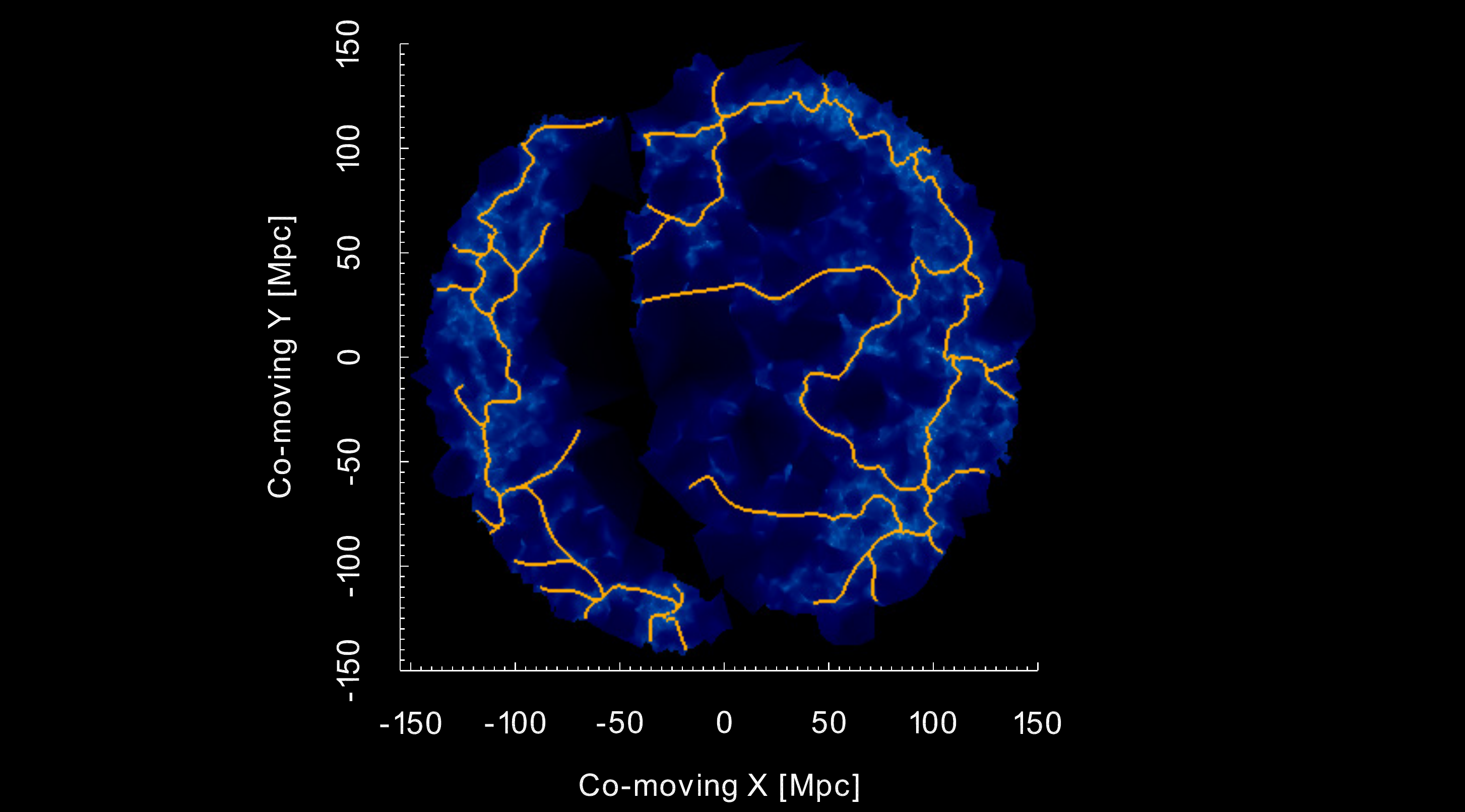}
\caption{Same as Fig.~\ref{fig:disperse01} for the redshift slices 0.0225 $\leq z <$ 0.0275 (left-hand column) and 0.0275 $\leq z <$ 0.00325 (right-hand column).}
\label{fig:disperse45}
\end{figure*}

% Slice 6 and 7
\begin{figure*}
\centering
\includegraphics[scale=.45, trim={5cm 0cm 7cm 0cm}, clip]{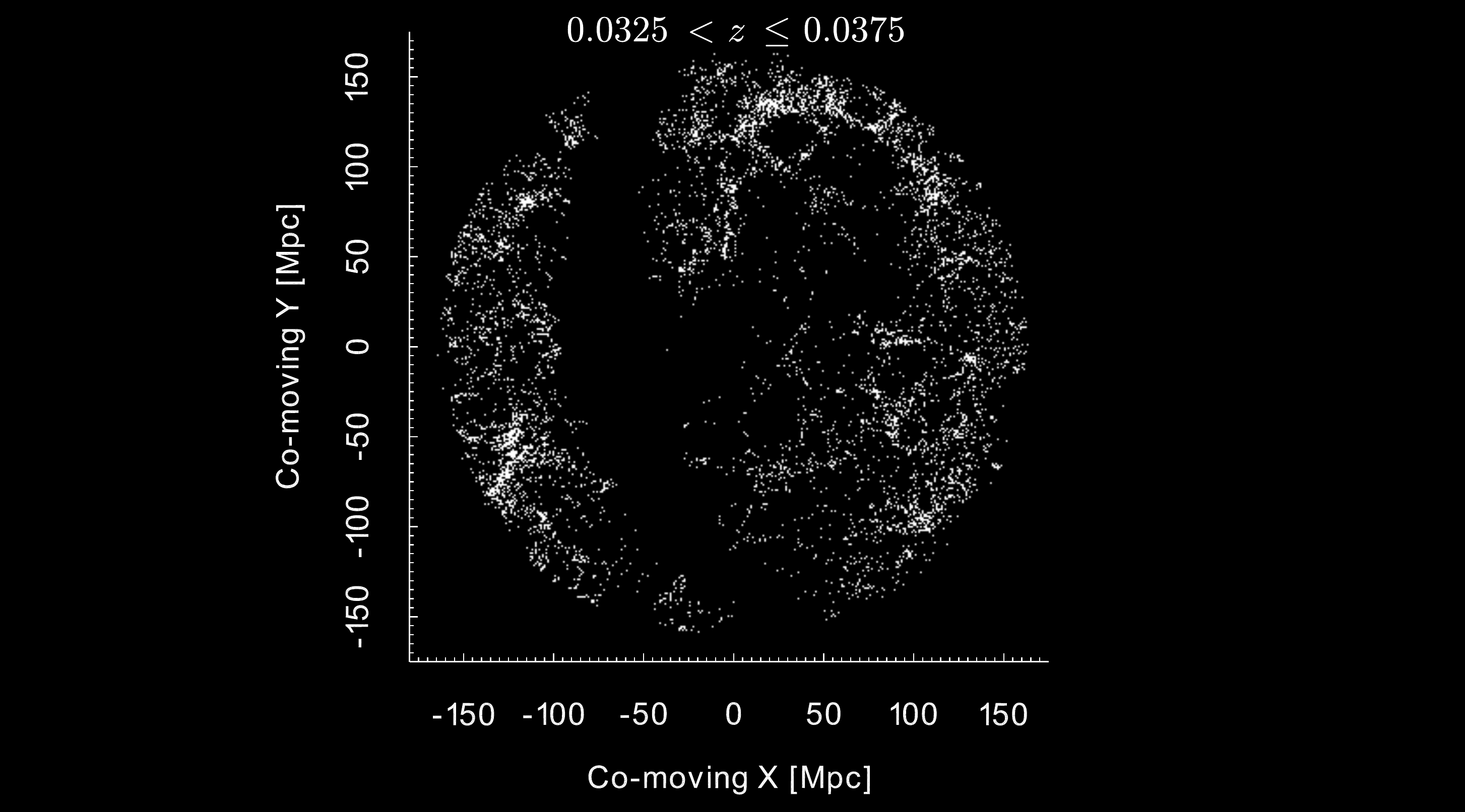}
\includegraphics[scale=.45, trim={5cm 0cm 7cm 0cm}, clip]{Galaxies_s7b-eps-converted-to}

\includegraphics[scale=.45, trim={5cm 0cm 7cm 0cm}, clip]{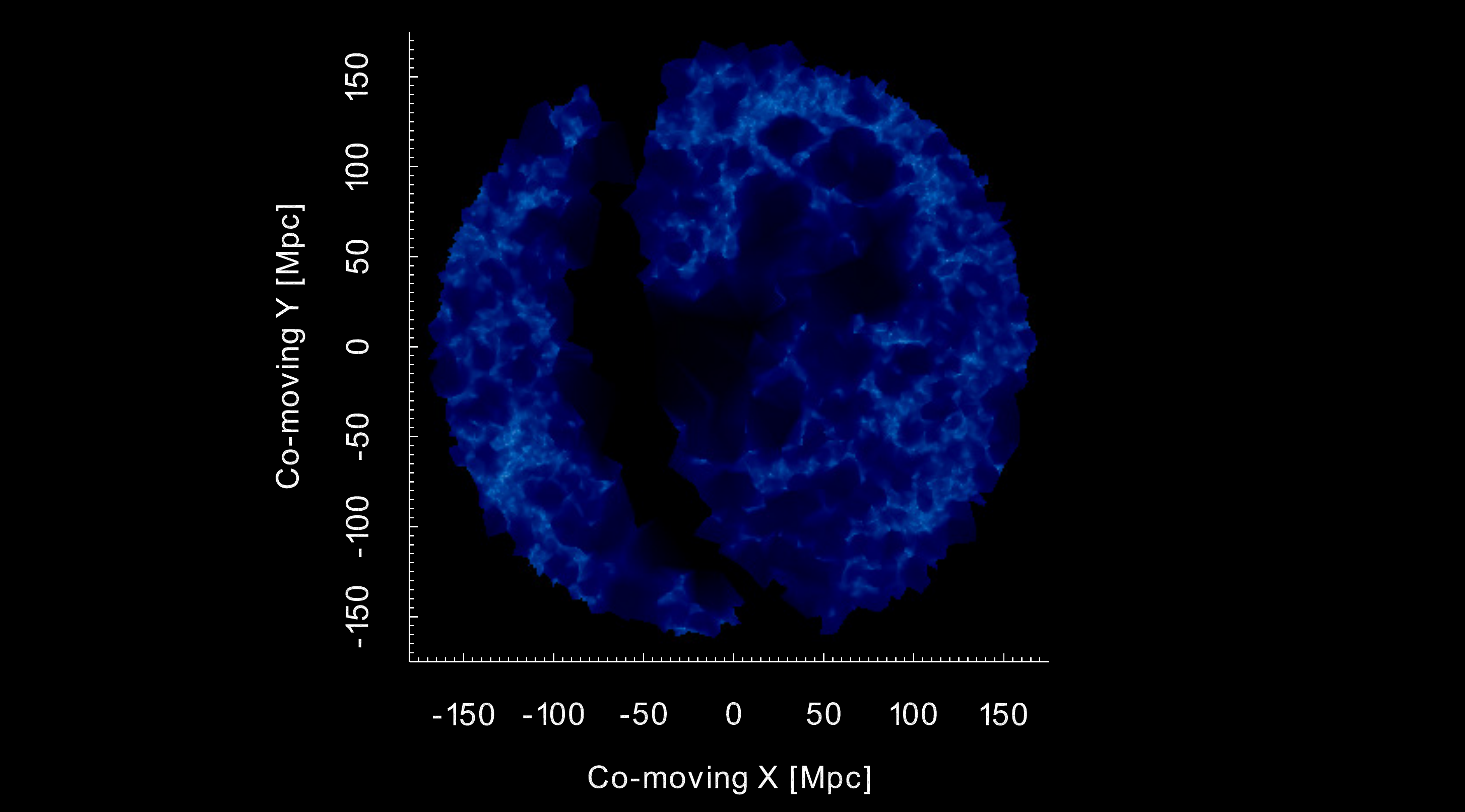}
\includegraphics[scale=.45, trim={5cm 0cm 7cm 0cm}, clip]{Delauany_tessellation_s7b-eps-converted-to}

\includegraphics[scale=.45, trim={5cm 0cm 7cm 0cm}, clip]{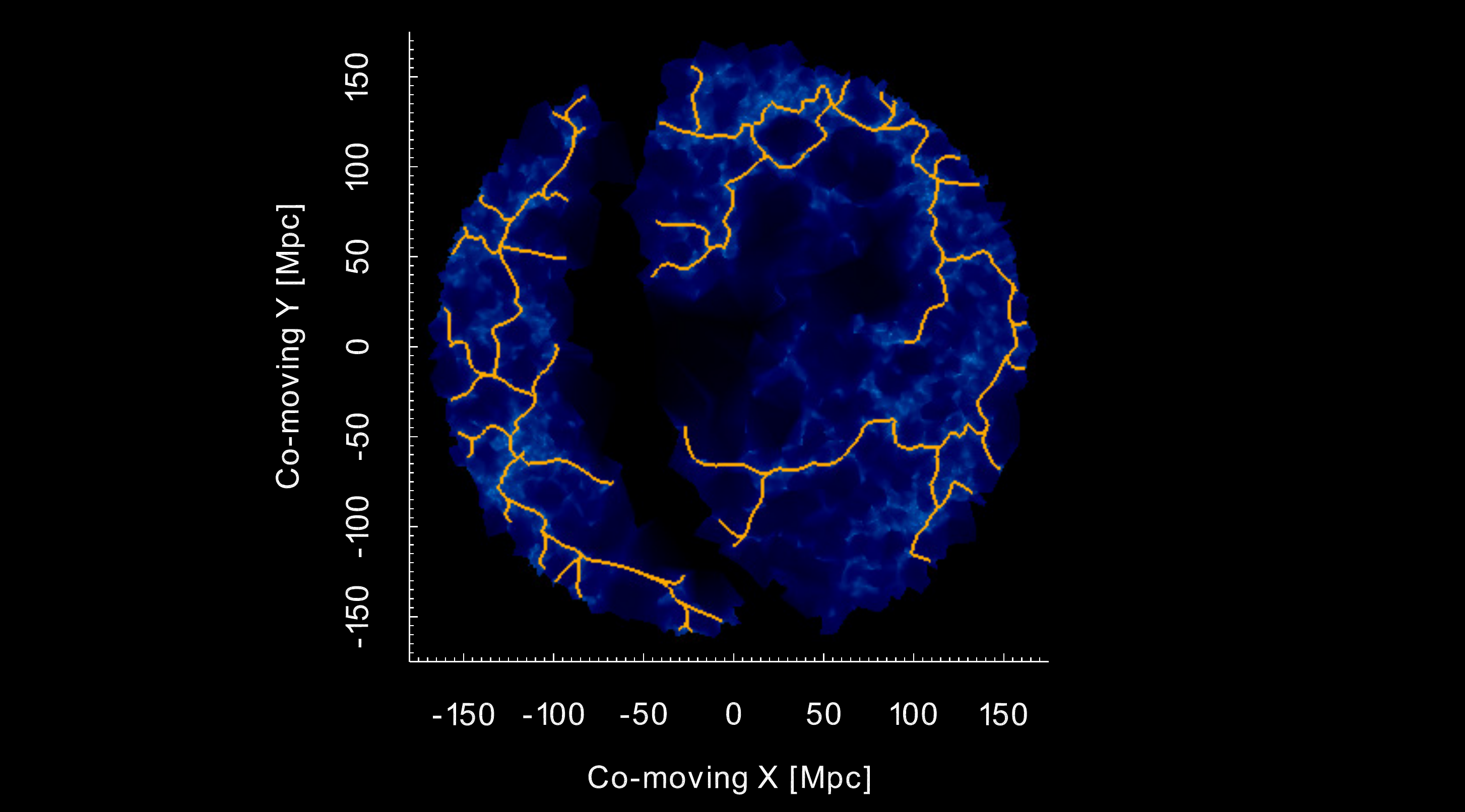}
\includegraphics[scale=.45, trim={5cm 0cm 7cm 0cm}, clip]{Filament_backbones_s7b-eps-converted-to}
\caption{Same as Fig.~\ref{fig:disperse01} for the redshift slices 0.0325 $\leq z <$ 0.0375 (left-hand column) and 0.0375 $\leq z <$ 0.0425 (right-hand column).}
\label{fig:disperse67}
\end{figure*}

\section{Comparison of two- and three-dimensional filaments found with DisPerSE}
\label{appen:disperse_compare}
In this appendix, we use the redshift slice 0.0375 $< z \leq$ 0.0425 to compare the filaments found with DisPerSE in two and three dimensions. This is the closest redshift slice in this work to the median redshift of 6dFGS and provides the optimal galaxy number density for classifying large-scale structure. 

Filaments found in two-dimensional data are described in Section \ref{fil_6dFGS} and Appendix \ref{appen:disperse}. The filament backbones delineated in this way show good agreement with the visible cosmic web components and removes the dependence of falsely identifying spurious filaments in the $z$-direction from galaxies with large peculiar velocities. 

\begin{figure}
\centering
\includegraphics[scale=.35, trim={6.5cm 0cm 8cm 0cm}, clip]{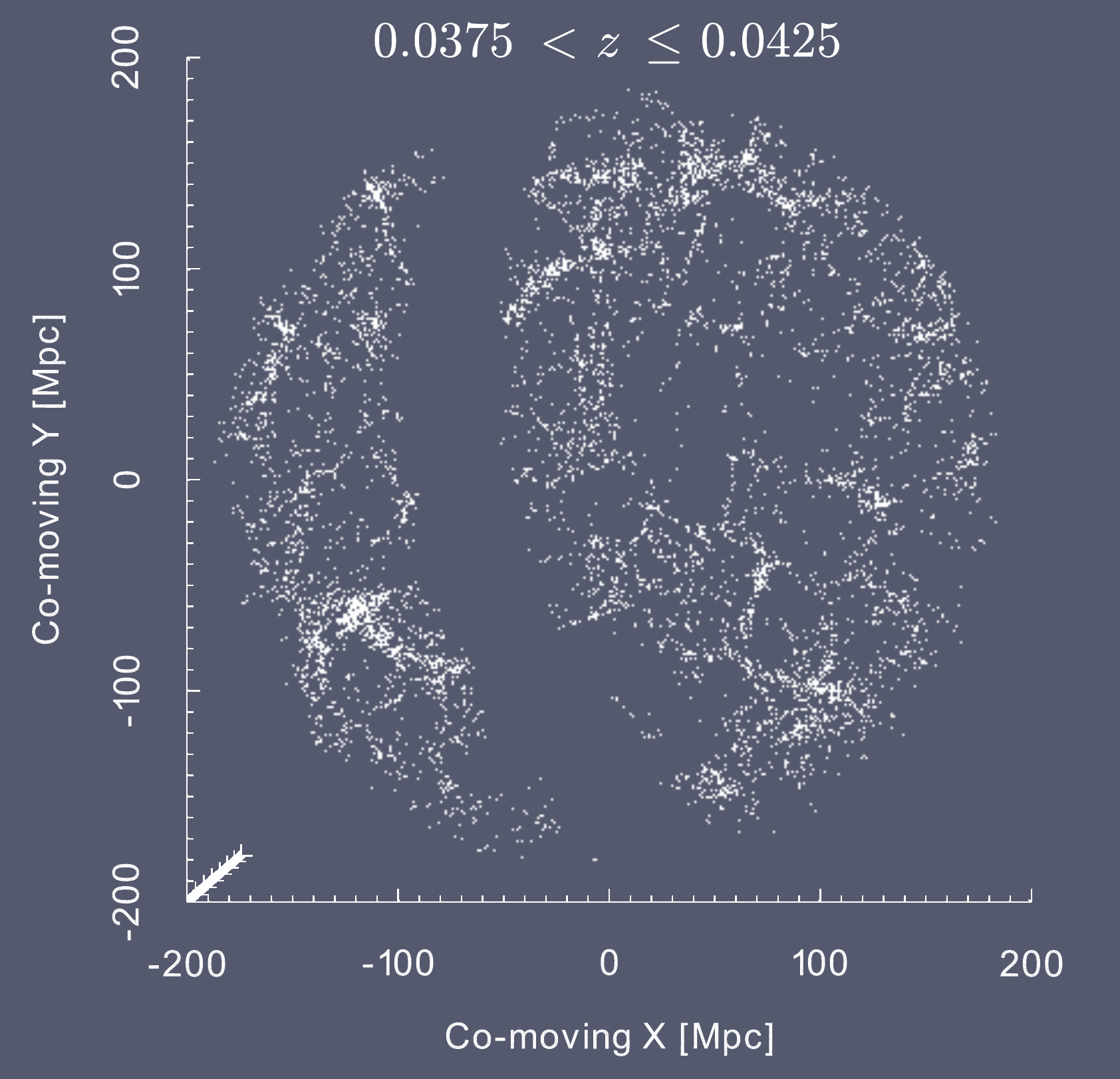}

\includegraphics[scale=.35, trim={6.5cm 0cm 8cm 0cm}, clip]{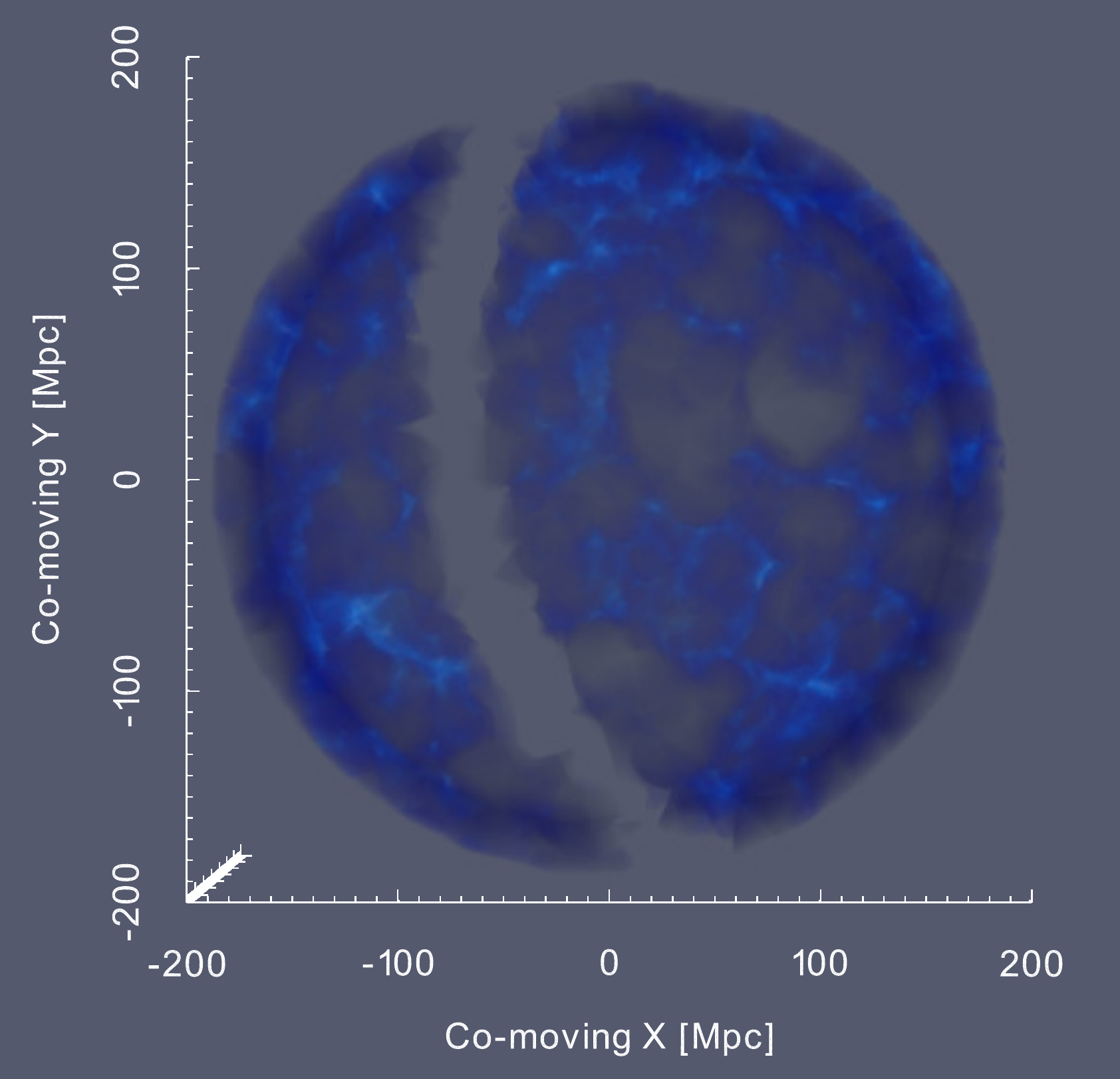}

\includegraphics[scale=.35, trim={6.5cm 0cm 8cm 0cm}, clip]{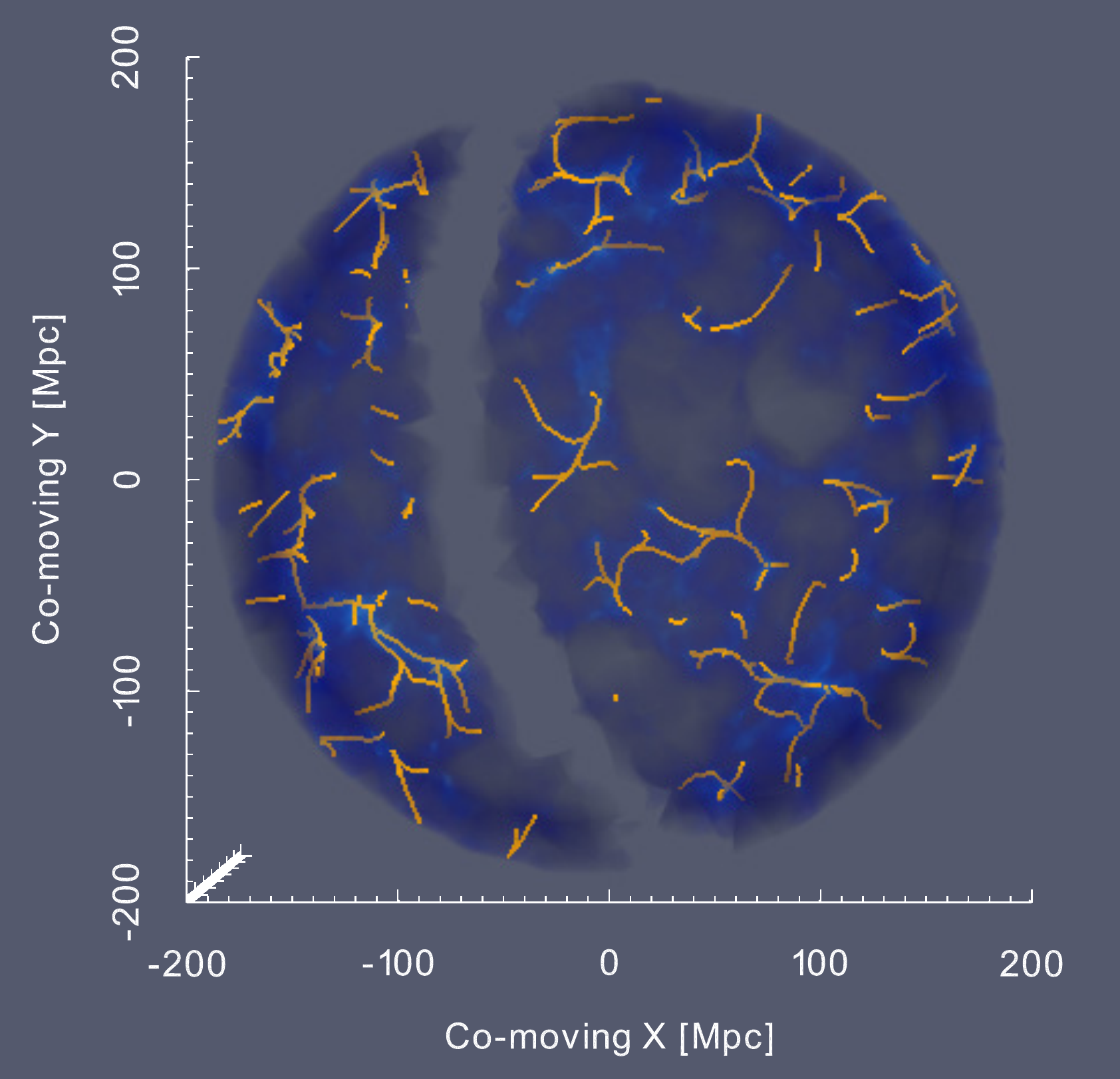}
\caption{The main steps of DisPerSE delineating filament backbones in three dimensions showing the results within 0.0375 $\leq z <$ 0.0425. The top panel shows all 6dFGS galaxies in the redshift range. The middle panel shows the corresponding Delaunay tessellation coloured by local density field with darker colours representing underdense regions and lighter colours highlighting the overdensities. The Delaunay tessellation in three dimensions is a volume compared to a surface in the two-dimensional case (middle panel of each figure in Appendix \ref{appen:disperse}. The bottom panel overlays the filament backbones (orange lines) found by DisPerSE with persistence ratios $\sigma_{\rm pr}$ = 4.0 and 4.8 for galaxies above and below the galactic plane. This sequence is the three-dimensional analogy of the right-hand panel of Fig.~\ref{fig:disperse67}. Even though there are similar topological features in the two- and three-dimensional tessellations, the filaments found from the tessellations are significantly different. The filaments found in two dimensions (Fig.~\ref{fig:disperse67}) trace the density ridges well but, the filaments in the three dimensions (bottom panel of this figure) appear unnecessarily complex and do not trace the thin features of the cosmic web.}
\label{fig:disperse_3D}
\end{figure}

To delineate filaments in three dimensions, an analogues method for the two-dimensional case is used. The galaxies are split into samples above and below the galactic plane, a Delaunay tessellation is produced from the three-dimensional galaxy positions and filaments are extracted as the thin topologically significant features. This procedure is applied to all galaxies in 6dFGS within the HIPASS volume and Fig.~\ref{fig:disperse_3D} shows these steps for the galaxies, tessellation and filaments that exist between 0.0375 $< z \leq$ 0.0425. Fig.~\ref{fig:disperse_3D} is the three-dimensional analogue of Fig.~\ref{fig:disperse67} (right-hand panel). Filaments in both cases were extracted with the same persistence ratios ($\sigma_{\rm pr}$ = 4.0 and 4.8 for galaxies with $b$ $>$ 10$^{\circ}$ and $b$ $<$ $-$10$^{\circ}$) and can therefore be directly compared to each other.

% 2D vs 3D

As the tessellation in Fig.~\ref{fig:disperse67} is two-dimensional and the tessellation in Fig.~\ref{fig:disperse_3D} is three-dimensional, we would not expect a one-to-one match of the topological features. However, they are a thin slice of the overall data and we expect the significant topological features to prevail in both tessellations. There is good agreement between the tessellations. For example, the high-density ridges centred at ($x,y$) $\approx$ ($-$100, $-$80), (80, 150) and the large underdense region centred at ($x,y$) $\approx$ (70, 50) appear in both the two- and three-dimensional tessellations. 

The filaments delineated in the two-dimensional tessellation (Fig.~\ref{fig:disperse67} bottom-right panel) clearly trace the thin density ridges that follow the network of the cosmic web. The filaments delineated in the three-dimensional tessellation (Fig.~\ref{fig:disperse_3D} bottom panel) are in the general location of the thin density ridges but they are a poor match to the network of the cosmic web. There are many more filaments found in the three-dimensional case and they appear to unnecessarily twist and turn. The three-dimensional filaments appear to preferentially circumscribe voids rather than trace the thin density ridges of the tessellation. While this may be an attempt at following the three-dimensional geometry of filaments, it defines numerous filaments in unrealistic regions and the reproduction of the cosmic web is unsatisfactory. Therefore, the filaments found in three dimensions do not match the two-dimensional filaments defined in the same volume.

\begin{figure*}
\includegraphics[scale=0.53]{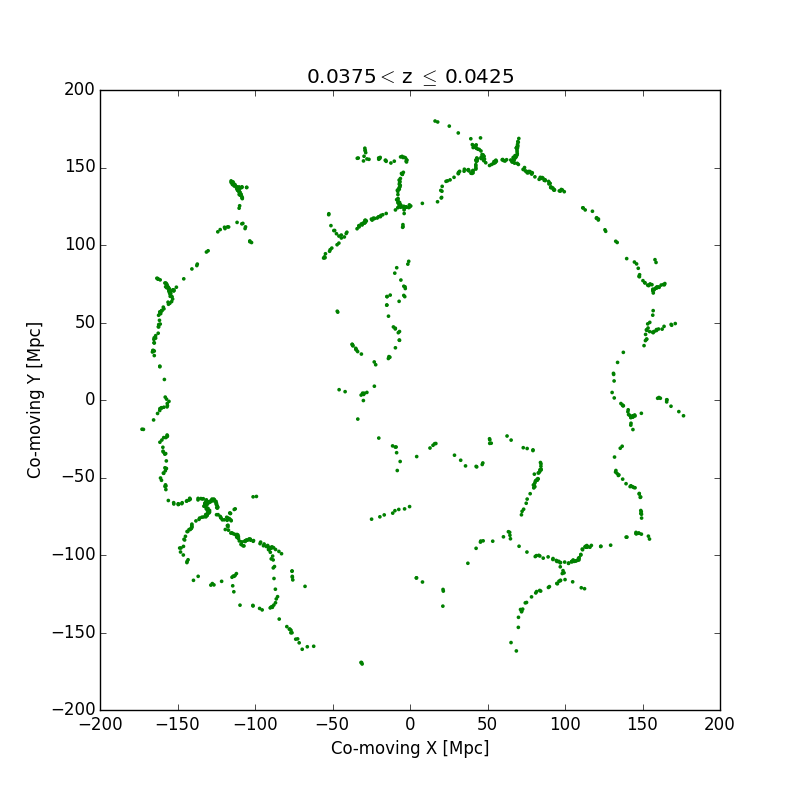}
 
\includegraphics[scale=0.53]{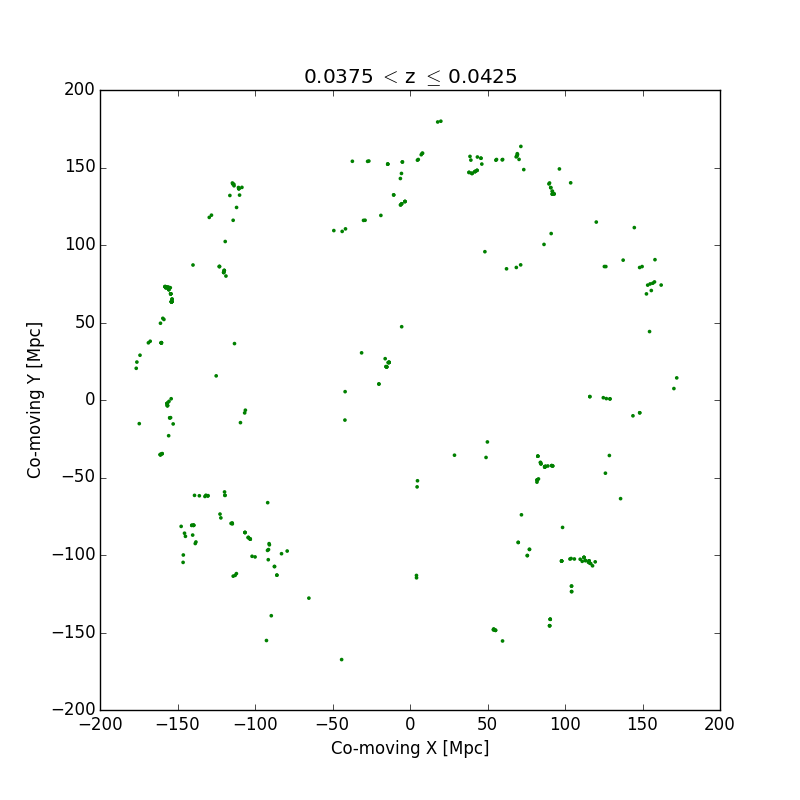}

\caption{A comparison of the galaxies (green points) contained in the near-filament samples produced in two (top panel) and three (bottom panel) dimensions. The two-dimensional sample consists of 1\,011 galaxies and was defined as galaxies within 0.7 Mpc of a filament backbone and a projected density of $\Sigma_{5}$ $<$ 3 galaxies Mpc$^{-2}$. The three-dimensional sample consists of 735 galaxies and was defined as galaxies within 0.7 Mpc with no (projected) density restriction. The two-dimensional near-filament sample produces filamentary structures while the three-dimensional near-filament sample does not. We expect the main geometry of the samples to match as this is a comparison in a thin redshift slice. The three-dimensional near-filament sample does not produce a satisfactory sample of filament galaxies and we use the two-dimensional sample definition in this work.}
\label{fig:2D_3D_gals}
\end{figure*}

We create \textit{near-filament} samples from the filaments backbones to compare which galaxies would be included in this analysis using the two- and three-dimensional filament delineation (Fig.~\ref{fig:2D_3D_gals}). The two-dimensional sample is described in Section \ref{samples} with galaxies being included if they are within 0.7 Mpc of filament backbones and a projected density of $\Sigma_{5}$ $<$ 3 galaxies Mpc$^{-2}$. The three-dimensional sample includes galaxies within 0.7 Mpc of filaments with no (projected) density restriction, making the number of galaxies included in this sample an upper limit. There are 1\,011 and 735 galaxies in the two- and three-dimensional near-filament samples. The two-dimensional sample outlines the filamentary structures in that redshift range, while the three-dimensional does not coherently trace the filamentary structures. Analogous to the tessellations, we do not expect the galaxy samples to exactly match in number or geometry. However, we are comparing a thin redshift slice and expect the main geometry of the samples to match. This is a reasonable expectation as the width of the redshift slice includes galaxies within $\pm$ 750 km s$^{-1}$, a velocity spread of the order of the Virgo clusters velocity dispersion. In this comparison, we find that the galaxy distribution of the near-filament samples in three dimensions does not reproduce the filaments of the cosmic web or match the main geometry of the two-dimensional near-filament sample. 

%% 2D and 3D filament galaxies

As we have found a systematic and reliable technique for delineating filaments in the nearby Universe (Section \ref{fil_6dFGS}), it is beyond the scope of this work to fully determine the cause of the discrepancies between delineating filaments in two and three dimensions. We believe that the following two factors are likely to be responsible for the discrepancies; (i) the peculiar velocities of galaxies in the nearby Universe are too high to coherently trace the three-dimensional geometry of filaments and (ii) the requirement for filaments in DisPerSE to circumscribe voids over powers tracing the thin density ridges. Both of these factors can explain why the three-dimensional delineation of filaments with DisPerSE was found to be unsatisfactory and we will determine the cause in a future study. 

The technique used in this work to delineate filaments in two dimensions is analogous to the technique used by \citet{Chen2015a, Chen2015b} who split SDSS into thin redshift slices (same thickness of $\delta z$ = 0.005 used in this work) to delineate filaments in two dimensions and study the optical properties of filament galaxies.

%%%%%%%%%%%%%%%%%%%%%%%%%%%%%%%%%%%%%%%%%%%%%%%%%%

% Don't change these lines
\bsp	% typesetting comment
\label{lastpage}
\end{document}